\documentclass[transmag]{IEEEtran}
\usepackage{latexsym}
\usepackage{graphicx}
\usepackage{amsfonts,amssymb,amsmath}
\usepackage{hyperref}
\def\BibTeX{{\rm B\kern-.05em{\sc i\kern-.025em b}\kern-.08em T\kern-.1667em\lower.7ex\hbox{E}\kern-.125emX}}

\usepackage[T1]{fontenc}
\usepackage{cite}

\graphicspath{../graphics/}
\DeclareGraphicsExtensions{.pdf,.jpeg,.png}

\usepackage[cmintegrals]{newtxmath}

\usepackage{algorithmic}
\usepackage{array}

\makeatletter
\let\MYcaption\@makecaption
\makeatother

\usepackage[font=footnotesize]{subcaption}

\makeatletter
\let\@makecaption\MYcaption
\makeatother

\usepackage{url} 

\usepackage{textcomp}
\usepackage{xcolor}
\usepackage{siunitx}
\usepackage{booktabs}
\usepackage{multirow}
\usepackage{arydshln}
\usepackage{graphbox}
\usepackage{paralist}
\usepackage[]{todonotes}
\usepackage{afterpage}
\usepackage{balance}

\DeclareMathOperator*{\argmax}{argmax} 

\definecolor{mod}{RGB}{0,0,155}

\begin{document}

\title{Multi-UAV Path Planning for Wireless Data Harvesting with Deep Reinforcement Learning}

\author{Harald~Bayerlein,~\IEEEmembership{Student~Member,~IEEE,}
        Mirco~Theile,~\IEEEmembership{Student~Member,~IEEE,}\\
        Marco~Caccamo,~\IEEEmembership{Fellow,~IEEE,}
        and~David~Gesbert,~\IEEEmembership{Fellow,~IEEE}

\thanks{H. Bayerlein and D. Gesbert were partially supported by the French government, through the 3IA Côte d’Azur project number ANR-19-P3IA-0002, as well as by the TSN CARNOT Institute under project Robots4IoT. M.~Caccamo was supported by an Alexander von Humboldt Professorship endowed by the German Federal Ministry of Education and Research. This article was presented in part at IEEE GLOBECOM 2020 \cite{Bayerlein2020}. The code for this work is available under \url{https://github.com/hbayerlein/uav_data_harvesting}. \textit{(Corresponding author: Harald Bayerlein)}}%

\thanks{H. Bayerlein and D. Gesbert are with the Communication Systems Department, EURECOM, Sophia Antipolis, France, \{harald.bayerlein, david.gesbert\}@eurecom.fr.}%

\thanks{M. Theile and M. Caccamo  are  with  the TUM Department of Mechanical Engineering, Technical University of Munich, Germany, \{mirco.theile, mcaccamo\}@tum.de.}}%

\IEEEtitleabstractindextext{
\begin{abstract}
Harvesting data from distributed Internet of Things (IoT) devices with multiple autonomous unmanned aerial vehicles (UAVs) is a challenging problem requiring flexible path planning methods. We propose a multi-agent reinforcement learning (MARL) approach that, in contrast to previous work, can adapt to profound changes in the scenario parameters defining the data harvesting mission, such as the number of deployed UAVs, number, position and data amount of IoT devices, or the maximum flying time, without the need to perform expensive recomputations or relearn control policies. We formulate the path planning problem for a cooperative, non-communicating, and homogeneous team of UAVs tasked with maximizing collected data from distributed IoT sensor nodes subject to flying time and collision avoidance constraints. The path planning problem is translated into a decentralized partially observable Markov decision process (Dec-POMDP), which we solve through a deep reinforcement learning (DRL) approach, approximating the optimal UAV control policy without prior knowledge of the challenging wireless channel characteristics in dense urban environments. By exploiting a combination of centered global and local map representations of the environment that are fed into convolutional layers of the agents, we show that our proposed network architecture enables the agents to cooperate effectively by carefully dividing the data collection task among themselves, adapt to large complex environments and state spaces, and make movement decisions that balance data collection goals, flight-time efficiency, and navigation constraints. Finally, learning a control policy that generalizes over the scenario parameter space enables us to analyze the influence of individual parameters on collection performance and provide some intuition about system-level benefits.
\end{abstract}

\begin{IEEEkeywords}
Internet of Things (IoT), map-based planning, multi-agent reinforcement learning (MARL), trajectory planning, unmanned aerial vehicle (UAV).
\end{IEEEkeywords}
}

\maketitle

\section{Introduction}

Autonomous unmanned aerial vehicles (UAVs) are not only envisioned as passive cellular-connected users of telecommunication networks but also as active connectivity enablers \cite{Zeng2019}. Their fast and flexible deployment makes them especially useful in situations where terrestrial infrastructure is overwhelmed or destroyed, e.g. in disaster and search-and-rescue situations \cite{Namuduri2017}, or where fixed coverage is in any way lacking. UAVs have shown particular promise in collecting data from distributed Internet of Things (IoT) sensor nodes. For instance, IoT operators can deploy UAV data harvesters in the absence of otherwise expensive cellular infrastructure nearby. Another reason is the throughput efficiency benefits related to having UAVs that describe a flight pattern that brings them close to the IoT devices. As an example in the context of infrastructure maintenance and preserving structural integrity, Hitachi is already commercially deploying partially autonomous UAVs that collect data from IoT sensors embedded in large structures, such as the San Juanico and Agas-Agas Bridges in the Philippines \cite{Minevich2020}. Research into UAV-aided data collection from IoT devices or wireless sensors include the works \cite{Cui2019,Pan2021,Tang2020,Zhang2020,Liu2021}, with \cite{Yi2020,Wu2020,Hu2020,AbdElmagid2019} concentrating on minimizing the age of information of the collected data. Additional coverage of past related work is offered in the next section.

In this work, we focus on controlling a team of UAVs, consisting of a variable number of identical drones tasked with collecting varying amounts of data from a variable number of stationary IoT sensor devices at variable locations in an urban environment. This imposes challenging constraints on the trajectory design for autonomous UAVs. In addition, the limited on-board battery energy density restricts mission duration for quadcopter drones severely. At the same time, the complex urban environment poses challenges in obstacle avoidance and adherence to regulatory no-fly zones (NFZs). Additionally, the wireless communication channel is characterized by random signal blocking events due to alternating between line-of-sight (LoS) and non-line-of-sight (NLoS) links. We believe this work is the first to address multi-UAV path planning where learned control policies are generalized over a wide scenario parameter space and can be directly applied when scenario parameters change without the need for retraining.

While some challenges to real-world deep reinforcement learning (DRL) such as limited training samples, safety and lack of explainable actions remain, DRL offers the opportunity to balance challenges and data collection goals for complex environments in a straightforward way by combining them in the reward function. Another reason for the popularity of the DRL paradigm in this context is the computational efficiency of DRL inference. DRL is also one of the few methods that allows us to tackle the complex task directly, given that UAV control and deployment in communication scenarios are generally non-convex optimization problems \cite{HuChen2020, Li2019, LiuXiao2019, Zeng2019, Shakeri2019, Saad2020}, and proven to be NP-hard in many instances \cite{LiuXiao2019, Zeng2019, Shakeri2019}. These advantages of DRL also hold for other UAV path planning instances, such as coverage path planning \cite{Theile2020}, a classical robotics problem where the UAV's goal is to cover all points inside an area of interest. The equivalence of these path planning problems and the connection between the often disjoint research areas is highlighted in \cite{Theile2021}.

\subsection{Related Work}
\label{subsec:related_work}

A survey that spans the various application areas for multi-UAV systems from a cyber-physical perspective is provided in \cite{Shakeri2019}. The general challenges and opportunities of UAV communications are summarized in publications by Zeng \textit{et al.} \cite{Zeng2019} and Saad \textit{et al.} \cite{Saad2020}, which both include data collection from IoT devices. This specific scenario is also included in \cite{Ullah2020} and \cite{Lahmeri2020}, surveys that comprise information on the classification of UAV communication applications with a focus on DRL methods.

Path planning for UAVs providing some form of communication services or collecting data has been studied extensively, including numerous approaches based on reinforcement learning (RL). However, it is crucial to note that the majority of previous works concentrates on only finding the optimal trajectory solution for one set of scenario parameters at a time, requiring full or partial retraining if the scenario changes. In contrast, our approach aims to train and generalize over a large scenario parameter space directly, finding efficient solutions without the need for lengthy retraining, but also increasing the complexity of the path planning problem significantly.

Many existing RL approaches also only focus on single-UAV scenarios. An early proposal given in \cite{Bayerlein2018spawc} to use (deep) RL in a related scenario where a single UAV base station serves ground users shows the advantages of using a deep Q-network (DQN) over table-based Q-learning, while not making any explicit assumptions about the environment at the price of long training time. The authors in \cite{Cui2019} only investigate table-based Q-learning for UAV data collection. A particular variety of IoT data collection is the one tackled in \cite{Yi2020}, where the authors propose a DQN-based solution to minimize the age of information of data collected from sensors. In contrast to our approach, the mentioned approaches are set in much simpler environments and agents have to undergo computationally expensive retraining when scenario parameters change.

Multi-UAV path planning for serving ground users employing table-based Q-learning is investigated in \cite{LiuXiao2019}, based on a relatively complex 3-step algorithm consisting of grouping the users with a genetic algorithm, then deployment and movement design in two separated instances of Q-learning. The investigated optimization problem is proven to be NP-hard, with Q-learning being confirmed as a useful tool to solve it. Pan \textit{et al.} \cite{Pan2021} investigate an instance of multi-UAV data collection from sensor nodes formulated as a classical traveling salesman problem without modeling the communication phase between UAV and node. The UAVs' trajectories are designed with a genetic algorithm that uses some aspects of DRL, namely training a deep neural network and experience replay. In contrast to the multi-stage optimization algorithms in \cite{LiuXiao2019} and \cite{Pan2021}, our approach consists of a more straightforward end-to-end DRL approach that scales to large and complex environments, generalizing over varying scenario parameters. 

The combination of DRL and multi-UAV control has been studied previously in various scenarios. The authors in \cite{Wu2020} focus on trajectory design for minimizing the age of information of sensing data generated by multiple UAVs themselves where the data can be either transmitted to terrestrial base stations or mobile cellular devices. Their focus lies on balancing the UAV sensing and transmission protocol in an unobstructed environment for one set of scenario parameters at a time. Other MARL path planning approaches to minimize the age of information of collected data include \cite{Hu2020} and \cite{AbdElmagid2019}. In \cite{Venturini2020}, a swarm of UAVs on a target detection and tracking mission in an unknown environment is controlled through a distributed DQN approach. While the authors also use convolutional processing to feed map information to the agents, the map is initially unknown and has to be explored to detect the targets. The agents' goal is to learn transferable knowledge that enables adaptation to new scenarios with fast relearning, compared to our approach to learn a control policy that generalizes over scenario parameters and requires no relearning.

Hu \textit{et al.} \cite{HuChen2020} proposed a distributed multi-UAV meta-learning approach to control a group of drone base stations serving ground users with random uplink access demands. While meta-learning allows them to reduce the number of training episodes needed to adapt to a new unseen uplink demand scenario, several hundred are still required. Our approach focuses on training directly on random but observable scenario parameters within a given value range, therefore not requiring retraining to adapt. Due to the small and obstruction-less environment, no maps are required in \cite{HuChen2020} and navigation constraints are omitted by keeping the UAVs at dedicated altitudes. In \cite{Tang2020}, multi-agent deep Q-learning is used to optimize trajectories and resource assignment of UAVs that collect data from pre-defined clusters of IoT devices and provide power wirelessly to them. The focus here is on maximizing minimum throughput in a wirelessly powered network without a complex environment and navigation constraints, only for a single scenario at a time. Similarly, in \cite{Zhang2020} there is also a strong focus on the energy supply of IoT devices through backscatter communications when a team of UAVs collects their data. The authors propose a multi-agent approach that relies on the definition of ambiguous boundaries between clusters of sensors. The scenario is set in a simple, unobstructed environment, not requiring maps or adherence to multiple navigation constraints, but requiring retraining when scenario parameters change.

In \cite{Liu2020coverage}, a group of interconnected UAVs is tasked with providing long-term communication coverage to ground users cooperatively. While the authors also formulate a POMDP that they solve by a DRL variant, there is no need for map information or processing. The scenario is set in a simple environment without obstacles or other navigation constraints. This work was extended under the paradigm of mobile crowdsensing, where mobile devices are leveraged to collect data of common interest in \cite{Liu2021}. The authors proposed a heterogeneous multi-agent DRL algorithm collecting data simultaneously with ground and aerial vehicles in an environment with obstacles and charging stations. While in this work, the authors also suggest a convolutional neural network to exploit a map of the environment, the small grid world does not necessitate extensive map processing. Furthermore, they do not center the map on the agent's position, which is highly beneficial \cite{Bayerlein2020}. In contrast to our method, control policies have to be relearned entirely in a lengthy training process for both mentioned approaches when scenario and environmental parameters change.

\subsection{Contributions}

If DRL methods are to be applied in any real-world mission, the prohibitively high training data demand poses one of the most severe challenges \cite{DulacArnold2019}. This is exacerbated by the fact that even minor changes in the scenario, such as in the number or location of sensor devices in data collection missions, typically requires repeating the full training procedure of the DRL agent. This is the case for existing approaches such as \cite{Liu2020coverage,Liu2021, Tang2020, Zhang2020, Wu2020, Hu2020, AbdElmagid2019, Bayerlein2018spawc}. Other approaches to reduce the training data demand include meta-learning \cite{HuChen2020} and transfer learning \cite{DulacArnold2019}. To the best of our knowledge, this is the first work that addresses this problem in path planning for multi-UAV data harvesting by proposing a DRL method that is able to generalize over a large space of scenario parameters in complex urban environments without prior knowledge of wireless channel characteristics based on centered global-local map processing.

The main contributions of this paper are the following:
\begin{itemize}
    \item We formulate a flying time constrained multi-UAV path planning problem to maximize harvested data from IoT sensors. We consider its translation to a decentralized partially observable Markov decision process (Dec-POMDP) with full reward function description in large, complex, and realistic environments that include no-fly zones, buildings that block wireless links (some possible to be flown over, some not), and dedicated start/landing zones.
    \item To solve the Dec-POMDP under navigation constraints without any prior knowledge of the urban environment's challenging wireless propagation conditions, we employ deep multi-agent reinforcement learning with centralized learning and decentralized execution.
    \item We show the advantage in learning and adaptation efficiency to large maps and state spaces through a dual global-local map approach with map centering over more conventional scalar neural network input in a multi-UAV setting.
    \item As perhaps our most salient feature, our algorithm offers \textit{parameter generalization}, which means that the learned control policy can be reused over a wide array of scenario parameters, including the number of deployed UAVs, variable start positions, maximum flying times, and number, location and data amount of IoT sensor devices, without the need to restart the training procedure as typically required by existing DRL approaches.
    \item Learning a generalized control policy enables us to compare performance over a large scenario parameter space directly. We analyze the influence of individual parameters on collection performance and provide some intuition about system-level benefits.
\end{itemize}

\subsection{Organization}

The paper is organized as follows: Section \ref{sec:system} introduces the multi-UAV mobility and communication channel model, which is translated to an MDP in Section \ref{sec:mdp} and followed by a description of the proposed map preprocessing in Section \ref{sec:map} and multi-agent DRL learning approach in Section \ref{sec:marl}. Simulation results and their discussion follow in Section \ref{sec:simulation}, and we conclude the paper with a summary and outlook to future work in Section \ref{sec:conclusion}.

\section{System Model}
\label{sec:system}

In the following, we present the key models for the multi-UAV path planning problem. Note that some level of simplification is needed when modeling the robots' dynamics in order to enable the implementation of the RL approach. Our assumptions are explicit whenever suitable.

We consider a square grid world of size $M \times M \in \mathbb{N}^2$ with cell size $c$ and the set of all possible positions $\mathcal{M}$. Discretization of the environment is a necessary condition for our map-processing approach, however note that our method can be applied to any rectangular grid world. The environment contains $L$ designated start/landing positions given by the set
$$
\mathcal{L}=\left\{ \left[x^l_i,y^l_i\right]^{\operatorname{T}}, ~i=1,\ldots,L,  ~: \left[x^l_i,y^l_i\right]^{\operatorname{T}} \in \mathcal{M}\right\}
$$
and the combination of the $Z$ positions the UAVs cannot occupy is given by the set
$$
\mathcal{Z}=\left\{ \left[x^z_i,y^z_i\right]^{\operatorname{T}}, ~i=1,\ldots,Z,  ~: \left[x^z_i,y^z_i\right]^{\operatorname{T}} \in \mathcal{M}\right\}.
$$
This includes tall buildings which the UAVs can not fly over and regulatory no-fly zones (NFZ). The number of $B$ obstacles blocking wireless links are given by the set
$$
\mathcal{B}=\left\{ \left[x^b_i,y^b_i\right]^{\operatorname{T}}, ~i=1,\ldots,B,  ~: \left[x^b_i,y^b_i\right]^{\operatorname{T}} \in \mathcal{M}\right\},
$$
representing all buildings, also smaller ones that can be flown over. The lowercase letters $l, z, b$ indicate the coordinates of the respective set of environmental features $\mathcal{L}, \mathcal{Z}, \mathcal{B}$. An example of a grid world is depicted in Fig. \ref{fig:example}, where obstacles, NFZs, start/landing zone, and an example of a single UAV trajectory are marked as described in the attached legend in Tab. \ref{table:legend}.

\afterpage{
    \begin{figure}[t!]
        \centering
        \begin{subfigure}{0.5\columnwidth}
        \includegraphics[width=\textwidth]{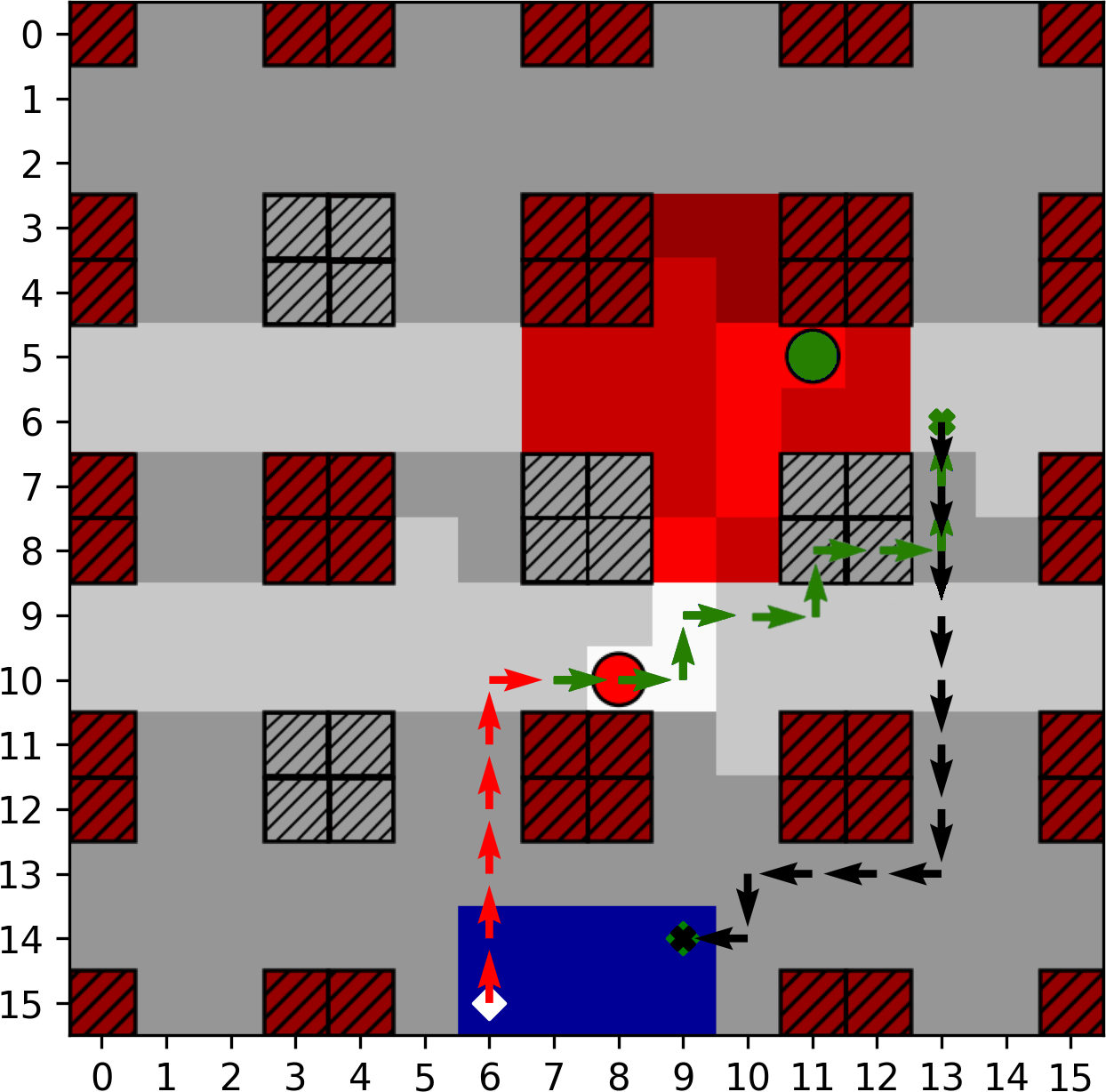}
        \end{subfigure}%
        \begin{subfigure}{0.01\columnwidth}
        \vspace{-4pt}
          $\left.
            \vcenter{\hbox{\vphantom{\includegraphics[width=49.1\textwidth]{graphics/mh1_globecom_new.png}}}}
          \right\rbrace M$
        \end{subfigure}
        \caption{Example of a single UAV collecting data from two IoT devices in an urban environment of size $M \times M$ with NFZs, a single start/landing zone, and buildings causing shadowing. Small buildings can be flown over and tall buildings act as navigation obstacles.}
        \label{fig:example}
    \end{figure}
    
    \begin{table}[h!]
    \center
    \small
    \begin{tabular*}{1.0\columnwidth}{lcl}
    \toprule[1.5pt]
    &Symbol & Description\\
    \midrule
    \multirow{6}{*}{\rotatebox[origin=c]{90}{\footnotesize{DQN Input}}}
    &\includegraphics[align=c,height=.3cm]{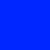} & Start and landing zone\\
    &\includegraphics[align=c,height=.3cm]{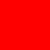} & Regulatory no-fly zone (NFZ)\\
    &\includegraphics[align=c,height=.3cm]{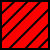} & Tall buildings* (UAVs cannot fly over)\\
    &\includegraphics[align=c,height=.3cm]{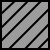} & Small buildings* (UAVs can fly over)\\
    &\includegraphics[align=c,height=.3cm]{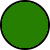} & IoT device\\
    &\includegraphics[align=c,height=.3cm]{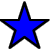} & Other agents\\\noalign{\vskip 3pt}
    & & *all buildings obstruct wireless links\\\midrule
    \multirow{5}{*}{\rotatebox[origin=c]{90}{\footnotesize{Visualization}}}
    &\includegraphics[align=c,height=.3cm]{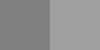} & Summation of building shadows\\
    &\includegraphics[align=c,height=.3cm]{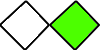} & Starting and landing positions during an episode\\
    &\includegraphics[align=c,height=.3cm]{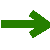} & UAV movement while comm. with  green device\\
    &\includegraphics[align=c,height=.3cm]{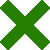} & Hovering while comm. with green device\\
    &\includegraphics[align=c,height=.3cm]{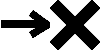} & Actions without comm. (all data collected)\\
    \bottomrule[1.5pt]
    \end{tabular*}
    \vspace{5pt}
    \caption{Legend for scenario plots.}
    \label{table:legend}
    \end{table}
}

\subsection{UAV Model}
\label{subsec:uav_model}

The set $\mathcal{I}$ of $I$ deployed UAVs moves within the limits of the grid world $\mathcal{M}$. The state of the $i$-th UAV is described through its:
\begin{itemize}
    \item position $\mathbf{p}_i(t) = [x_i(t), y_i(t), z_i(t)]^{\operatorname{T}} \in \mathbb{R}^3$ with altitude $z_i(t) \in \{0, h\}$, either at ground level or in constant altitude $h$;
    \item operational status $\phi_i(t) \in \{0,1\}$, either inactive or active;
    \item battery energy level $b_i(t) \in \mathbb{N}$.
\end{itemize}
Note that the assumption of all UAVs sharing the same flying altitude is not too restrictive and that our method allows each UAV to fly at a different altitude as long as it remains constant throughout the mission. The UAV agent's altitude can be made observable by simply adding it to the observation space along the flying time. This work only tackles 2D trajectory optimization, as the environment is dominated by high-rise buildings that would require long climbing phases to be overflown. The mission time limited by the UAVs' on-board batteries restricts the effectiveness of 3D control for the data collection performance given that climbing flight consumes more energy \cite{LiuSengupta2017} and that the UAVs needs to land at ground level at the end of the mission. The data collection mission is over after $T \in \mathbb{N}$ mission time steps for all UAVs, where the time horizon is discretized into equal mission time slots $t \in [0, T]$ of length $\delta_t$ seconds.

The action space of each UAV is defined as
\begin{equation}
    \mathcal{A} = 
    \Bigg\{
    \underbrace{\begin{bmatrix}0\\0\\0\end{bmatrix}}_{\text{hover}},
    \underbrace{\begin{bmatrix}c\\0\\0\end{bmatrix}}_{\text{east}}, 
    \underbrace{\begin{bmatrix}0\\c\\0\end{bmatrix}}_{\text{north}},
    \underbrace{\begin{bmatrix}-c\\0\\0\end{bmatrix}}_{\text{west}},
    \underbrace{\begin{bmatrix}0\\-c\\0\end{bmatrix}}_{\text{south}},
    \underbrace{\begin{bmatrix}0\\0\\-h\end{bmatrix}}_{\text{land}}
    \Bigg\}.
\end{equation}
Each UAV's movement actions $\mathbf{a}_{i}(t) \in \tilde{\mathcal{A}}(\mathbf{p}_i(t))$ are limited to
\begin{equation}\label{eq:feasible_actions}
    \tilde{\mathcal{A}}(\mathbf{p}_i(t)) = 
    \begin{cases}
        \mathcal{A}, & \mathbf{p}_i(t) \in \mathcal{L}\\
        \mathcal{A} \setminus [0,0,-h]^\text{T}, & \text{otherwise},
    \end{cases}
\end{equation}
where $\tilde{\mathcal{A}}$ defines the set of feasible actions depending on the respective UAV's position, specifically that the landing action is only allowed if the UAV is in the landing zone.

The distance the UAV travels within one time slot is equivalent to the cell size $c$. Mission time slots are chosen sufficiently small so that each UAV's velocity $v_i(t)$ can be considered to remain constant in one time slot. The UAVs are limited to moving with horizontal velocity $V=c / \delta_t$ or standing still, i.e. $v_i(t)\in \{0, V\}$ for all $t \in [0, T]$. Each UAV's position evolves according to the motion model given by
\begin{equation}
     \mathbf{p}_i(t+1) = 
     \begin{cases}
     \mathbf{p}_i(t) + \mathbf{a}_{i}(t), & \phi_i(t)=1\\
     \mathbf{p}_i(t), & \text{otherwise},
     \end{cases}
\end{equation}
keeping the UAV stationary if inactive. The evolution of the operational status $\phi_i(t)$ of each UAV is given by
\begin{equation}\label{eq:operation_status}
    \phi_i(t+1) =
    \begin{cases}
    0, &\phantom{\lor~}\mathbf{a}_i(t) = [0,0,-h]^\text{T} \\
     &\lor~ \phi_i(t) = 0\\
    1, &\text{otherwise},\\
    \end{cases}
\end{equation}
where the operational status becomes inactive when the UAV has safely landed. The end of the data harvesting mission $T$ is defined as the time slot when all UAVs have reached their terminal state and are not actively operating anymore, i.e. the operational state is $ \phi_i(t) = 0$ for all UAVs.

The $i$-th UAV's battery content evolves according to 
\begin{equation}\label{eq:battery_evolve}
b_i(t+1) = 
\begin{cases}
    b_i(t) - 1,  & \phi_i(t) = 1\\
    b_i(t) , &\text{otherwise,}\\
\end{cases}
\end{equation}
assuming a constant energy consumption while the UAV is operating and zero energy consumption when operation has terminated. This is a simplification justified by the fact that power consumption for small quadcopter UAVs is dominated by the hovering component. Using the model from \cite{LiuSengupta2017}, the ratio between the additional power necessary for horizontal flight at $10\si{\meter/\second}$ and just hovering could be roughly estimated as $30\si{\watt}/310\si{\watt} \approx 10\%$, which is negligible. Considering power consumption of on-board computation and communication hardware which does not differ between flight and hovering, the overall difference becomes even smaller. In the following, we will refer to the battery content as remaining flying time, as it is directly equivalent.

The overall multi-UAV mobility model is restricted by the following constraints:
\begin{subequations}\label{eq:constraints}
\begin{align}
    \begin{split}\label{eq:collision_constraint}
        \mathbf{p}_{i}(t) \neq \mathbf{p}_{j}(t) \lor \phi_{j}(t) = 0,\quad &\forall i,j \in \mathcal{I}, i\neq j, \forall t
    \end{split}\\
    \begin{split}\label{eq:nfz_constraint}
         \mathbf{p}_{i}(t) \notin \mathcal{Z}, \quad &\forall i \in \mathcal{I}, \forall t 
    \end{split}\\
    \begin{split}\label{eq:landing_constraint}
    b_i(t) \geq 0, \quad &\forall i \in \mathcal{I}, \forall t
    \end{split}\\
    \begin{split}\label{eq:starting_constraint}
    \mathbf{p}_{i}(0) \in \mathcal{L}\land z_i(0)=h, \quad &\forall i \in \mathcal{I}
    \end{split}\\
    \begin{split}\label{eq:operational_start}
    \phi_i(0)=1,  \quad &\forall i \in \mathcal{I}
    \end{split}
\end{align}
\end{subequations}
The constraint \eqref{eq:collision_constraint} describes collision avoidance among active UAVs with the exception that UAVs can land at the same location. \eqref{eq:nfz_constraint} forces the UAVs to avoid collisions with tall obstacles and prevents them from entering NFZs. The constraint \eqref{eq:landing_constraint} limits operation time of the drones, forcing UAVs to end their mission before their battery has run out. Since operation can only be concluded with the landing action as described in \eqref{eq:operation_status} and the landing action is only available in the landing zone as defined in \eqref{eq:feasible_actions}, the constraint \eqref{eq:landing_constraint} ensures that each UAV safely lands in the landing zone before their batteries are empty. The starting constraint \eqref{eq:starting_constraint} defines that the UAV start positions are in the start/landing zones and that their starting altitude is $h$, while  \eqref{eq:operational_start} ensures that the UAVs start in the active operational state.

\subsection{Communication Channel Model}
\label{subsec:comm_model}

\subsubsection{Link Performance Model}
As communication systems typically operate on a smaller timescale than the UAVs' mission planning system, we introduce the notion of communication time slots in addition to mission time slots. We partition each mission time slot $t \in [0, T]$ into a number of $\lambda \in \mathbb{N}$ communication time slots. The communication time index is then $n \in [0, N]$ with $N = \lambda T$. One communication time slot $n$ is of length $\delta_n = \delta_t/\lambda$ seconds. The number of communication time slots per mission time slot $\lambda$ is chosen sufficiently large so that the $i$-th UAV's position, which is interpolated linearly between $\mathbf{p}_i(t)$ and $\mathbf{p}_i(t+1)$, and the channel gain can be considered to stay constant within one communication time slot. 

The $k$-th IoT device is located on ground level at $\mathbf{u}_k = [x_k,y_k,0]^{\operatorname{T}} \in \mathbb{R}^3$ with $k \in \mathcal{K}$ where $|\mathcal{K}|=K$. Each IoT sensor has a finite amount of data $D_k(t) \in \mathbb{R}^{\operatorname{+}}$ that needs to be picked up over the whole mission time $t \in [0, T]$. The device data volume is set to an initial value at the start of the mission $D_k(t=0) = D_{k, init}$. The data volume of each IoT node evolves depending on the communication time index $n$ over the whole mission time, given by $D_k(n)$ with $n \in [0, N], N = \lambda T$.

We follow the same UAV-to-ground channel model as used in \cite{Bayerlein2020}. The communication links between UAVs and the $K$ IoT devices are modeled as LoS/NLoS point-to-point channels with log-distance path loss and shadow fading. The maximum achievable information rate at time $n$ for the $k$-th device is given by
\begin{equation}\label{eq:achieveable_rate}
    R^{\text{max}}_{i,k}(n) =  \log_2 \left( 1 + \text{SNR}_{i,k}(n) \right).
\end{equation}
Considering the amount of data available at the $k$-th device $D_k(n)$, the effective information rate is given as
\begin{equation}\label{eq:rate}
    R_{i,k}(n) = 
    \begin{cases}
    R^{\text{max}}_{i,k}(n), &D_k(n) \geq \delta_n R^{\text{max}}_{i,k}(n) \\
    D_k(n) / \delta_n, &\text{otherwise}.
    \end{cases}
\end{equation}
The SNR with transmit power $P_{i,k}$, white Gaussian noise power at the receiver $\sigma^2$, UAV-device distance $d_{i,k}$, path loss exponent $\alpha_e$ and $\eta_e \sim  \mathcal{N}(0,\,\sigma_e^{2})$ modeled as a Gaussian random variable, is defined as
\begin{equation}\label{eq:snr}
    \text{SNR}_{i,k}(n) = \frac{P_{i,k}}{\sigma^2} \cdot d_{i,k}(n)^{-\alpha_e} \cdot 10^{\eta_e/10}.
\end{equation}
Note that the urban environment with the set of obstacles $\mathcal{B}$ hindering free propagation causes a strong dependence of the propagation parameters on the $e \in \{\text{LoS, NLoS}\}$ condition and that \eqref{eq:snr} is the SNR averaged over small scale fading. We would also like to point out that our DQN-based trajectory planning approach is model-free and does therefore not rely on any specific channel model. While a more accurate and complex model could be directly used with our approach, the most important features for data collection missions of the urban channel, the dependence of SNR on $d_{i,k}$ and the $e \in \{\text{LoS, NLoS}\}$ condition, are already captured in \eqref{eq:snr}.

\subsubsection{Multiple Access Protocol}

The multiple access protocol is assumed to follow the standard time-division multiple access (TDMA) model when it comes to the communication between one single UAV and the various ground nodes. We further assume that the communication channel between the ground nodes and a given UAV operates on resource blocks (time-frequency slots) that are orthogonal to the channels linking the ground nodes and other UAVs, so that no inter-UAV interference exists in our model and inter-UAV synchronization is not necessary. Hence, the UAVs are similar to base stations that would be assigned orthogonal spectral resources. We also assume that IoT devices are operating in multi-band mode, hence are capable of simultaneously communicating with all UAVs on the set of all orthogonal frequencies. As a consequence, scheduling decisions are not part of the action space. The number of available orthogonal subchannels for UAV-to-ground communication is one of the variable scenario parameters and equivalent to the number of deployed UAVs.

Designing multiple access protocols for UAV networks is in itself a challenging research problem \cite{Hentati2020} due to high mobility of the nodes and fast changing link performance and is out of scope for this work. However, our proposed algorithm can in principle be integrated with existing solutions and does not rely on any specific channel model or multiple access protocol. While our model avoids and does not consider inter-UAV interference, we would like to point out that the behavior of the UAV agents that emerges naturally during the learning process of dividing the data collection task geographically, as illustrated in section \ref{subsec:urban50}, would mitigate the influence of interference on the trajectory planning decisions to some extent.

Our scheduling protocol is assumed to follow the max-rate rule: in each communication time slot $n \in [0, N]$, the sensor node $k \in [1, K]$ with the highest $\text{SNR}_{i, k}(n)$ with remaining data to be uploaded is picked by the scheduling algorithm. The TDMA constraint for the scheduling variable $q_{i, k}(n) \in \{0, 1\}$ is given by
\begin{equation}\label{eq:tdma_constraint}
    \sum_{k=1}^K q_{i,k}(n) \leq 1,~ n \in\left[0, N\right], \forall i \in \mathcal{I}.
\end{equation}
It follows that the $k$-th device's data volume evolves within one communication time slot according to
\begin{equation}
    D_k(n+1) = D_k(n) - \sum_{i=1}^{I}q_{i,k}(n) R_{i,k}(n)\delta_n.
\end{equation}

The achievable throughput for the $i$-th UAV for one mission time slot $t \in [0, T]$, comprised of $\lambda$ communication time slots, is the sum of rates achieved in the communication time slots $n \in [\lambda t, \lambda (t + 1) - 1]$  over $K$ sensor nodes. It depends on the UAV's operational status $\phi_i(t)$ and is given by
\begin{equation} \label{eq:throughput}
    C_{i}(t) = \phi_i(t) \sum_{n=\lambda t}^{\lambda (t + 1) - 1} \sum_{k=1}^K q_{i,k}(n) R_{i,k}(n)\delta_n.
\end{equation}

\subsection{Optimization Problem}
\label{subsec:optimization_problem}
Using the described UAV model in \ref{subsec:uav_model} and communication model in \ref{subsec:comm_model}, the central goal of the multi-UAV path planning problem is the maximization of throughput over the whole mission time and over all $I$ deployed UAVs while adhering to mobility constraints \eqref{eq:collision_constraint}-\eqref{eq:operational_start} and the scheduling constraint \eqref{eq:tdma_constraint}. The maximization problem is given by
\begin{align}
    \max_{\times_i \mathbf{a}_i(t)}~ \sum_{t = 0}^{T} \sum_{i = 1}^{I}  C_{i}(t).\label{eq:optimization_problem}\\
    \text{s.t.}\quad \eqref{eq:collision_constraint},\eqref{eq:nfz_constraint},\eqref{eq:landing_constraint},\eqref{eq:starting_constraint},\eqref{eq:operational_start},\eqref{eq:tdma_constraint}\nonumber
\end{align}
optimizing over joint actions $\times_i \mathbf{a}_i(t)$.

\section{Markov Decision Process (Dec-POMDP)}
\label{sec:mdp}

To address the aforementioned optimization problem, we translate it to a decentralized partially observable Markov decision process (Dec-POMDP) \cite{Oliehoek2016}, which is defined through the tuple $(\mathcal{S}, \mathcal{A}_\times, P, R, \Omega_\times, \mathcal{O}, \gamma)$. In the Dec-POMDP, $\mathcal{S}$ describes the state space, $\mathcal{A}_\times = \mathcal{A}^I$ the joint action space, and $P : \mathcal{S}\times\mathcal{A}_\times\times\mathcal{S}\mapsto\mathbb{R}$ the transition probability function. $R : \mathcal{S}\times\mathcal{A}\times\mathcal{S}\mapsto\mathbb{R}$ is the reward function mapping state, individual action, and next state to a real valued reward. The joint observation space is defined through $\Omega_\times = \Omega^I$ and $\mathcal{O}:\mathcal{S}\times \mathcal{I}\mapsto \Omega$ is the observation function mapping state and agents to one agent's individual observation. The discount factor $\gamma \in [0,1]$ controls the importance of long vs. short term rewards.

\subsection{State Space}
The state space of the multi-agent data collection problem consists of the environment information, the state of the agents, and the state of the devices. It is given as
\begin{align}
    \mathcal{S} = 
    &\underbrace{\mathcal{L}}_{\substack{\text{Landing}\\ \text{Zones}}}\times 
    \underbrace{\mathcal{Z}}_{\text{NFZs}}\times 
    \underbrace{\mathcal{B}}_{\text{Obstacles}}&\Big\}&\text{Environment}\nonumber \\
    &\times\underbrace{\mathbb{R}^{I\times 3}}_{\substack{\text{UAV}\\ \text{Positions}}}\times
    \underbrace{ \mathbb{N}^I}_{\substack{\text{Flying}\\ \text{Times}}}\times
    \underbrace{\mathbb{B}^{I}}_{\substack{\text{Operational}\\ \text{Status}}}&\Big\}&\text{Agents}\label{eq:state_space}\\
    &\times\underbrace{\mathbb{R}^{K\times3}}_{\substack{\text{Device}\\ \text{Positions}}}\times
    \underbrace{\mathbb{R}^{K}}_{\substack{\text{Device}\\ \text{Data}}}&\Big\}&\text{Devices}\nonumber
\end{align}
in which the elements $s(t) \in \mathcal{S}$ are
\begin{equation}
    s(t) = (\mathbf{M}, \{\mathbf{p}_i(t)\}, \{b_i(t)\}, \{\phi_i(t)\}, \{\mathbf{u}_k\}, \{D_k(t)\}),
\end{equation}
$\forall i \in \mathcal{I}$ and $\forall k \in \mathcal{K}$, in which $\mathbf{M}\in \mathbb{B}^{M\times M\times 3}$ is the tensor representation of the set of start/landing zones $\mathcal{L}$, obstacles and NFZs $\mathcal{Z}$, and obstacles only $\mathcal{B}$. The other elements of the tuple define positions, remaining flying times, and operational status of all agents, as well as positions and available data volume of all IoT devices.

\subsection{Safety Controller}
To enforce the collision avoidance constraint \eqref{eq:collision_constraint} and the NFZ and obstacle avoidance constraint \eqref{eq:nfz_constraint}, a safety controller is introduced into the system. Additionally, the safety controller enforces the limited action space excluding the \textit{landing} action when the respective agent is not in the landing zone as defined in \eqref{eq:feasible_actions}. The safety controller evaluates the action $\mathbf{a}_i(t)$ of agent $i$ and determines if it should be accepted or rejected. If rejected, the resulting safe action is the \textit{hovering} action. The safe action $\mathbf{a}_{s,i}(t)$ is thus defined as
\begin{equation}\label{eq:safe_action}
    \mathbf{a}_{s,i}(t) =
    \begin{cases}
    [0,0,0]^\text{T}, & \phantom{\lor~} \mathbf{p}_i(t) + \mathbf{a}_i(t) \in \mathcal{Z} \\
&\lor~ \mathbf{p}_i(t) + \mathbf{a}_i(t) = \mathbf{p}_j(t) \land \phi_j(t)=1,\\
&\phantom{\lor~}\forall j, j \neq i\\
&\lor~ \mathbf{a}_i(t) = [0,0,-h]^\text{T} ~\land~  \mathbf{p}_i(t)\notin \mathcal{L}\\
    ~\mathbf{a}_i(t),& \text{otherwise.}
    \end{cases}
\end{equation}
Without path planning capabilities, the safety controller cannot enforce the flying time and safe landing constraint in \eqref{eq:landing_constraint}. Therefore, we relax the hard constraint on flight time by adding a high penalty on not landing in time instead. In the simulation, a crashed agent, i.e. an agent with $b_i(t) < 0$, is defined as not operational.

\subsection{Reward Function}
\label{subsec:reward}
The reward function $R: \mathcal{S}\times\mathcal{A}\times\mathcal{S}\mapsto\mathbb{R}$ of the Dec-POMDP is comprised of the following elements:
\begin{equation}
    r_i(t) = \alpha \sum_{k\in\mathcal{K}}\Big(D_{k}(t+1)-D_{k}(t)\Big) + \beta_i(t) + \gamma_i(t) + \epsilon.
\end{equation}
The first term of the sum is a collective reward for the collected data from all devices by all agents within mission time \mbox{slot $t$}. It is parameterized through the data collection multiplier $\alpha$. This is the only part of the reward function that is shared among all agents. The second addend is an individual penalty when the safety controller rejects an action and given through
\begin{equation}
    \beta_i(t)=
\begin{cases}
\beta , & \mathbf{a}_i(t) \neq \mathbf{a}_{i,s}(t) \\ 
0,  & \text{otherwise}.\\
\end{cases}
\end{equation}
It is parameterized through the safety penalty $\beta$. The third term is the individual penalty for not landing in time given by
\begin{equation}
    \gamma_i(t)=
\begin{cases}
\gamma , & b_i(t+1) = 0 \land \mathbf{p}_i(t+1) = [\cdot, \cdot, h]^\text{T} \\ 
0,  & \text{otherwise}.\\
\end{cases}
\end{equation}
and parameterized through the crashing penalty $\gamma$. The last term is a constant movement penalty parameterized through $\epsilon$, which is supposed to incentivize the agents to reduce their flying time and prioritize efficient trajectories.

\section{Map-Processing and Observation Space}
\label{sec:map}
To aid the agents in interpreting the large state space given in \eqref{eq:state_space}, we implement two map processing steps. The first is centering the map around the agent's position, shown in \cite{Bayerlein2020} to significantly improve the agent's learning performance. This benefit is a consequence of neurons in the layer after the convolutional layers (compare Fig. \ref{fig:network}) corresponding to features \textit{relative} to the agent's position, rather than to \textit{absolute} positions if the map is not centered. This is advantageous as one agent's actions are solely based on its relative position to features, e.g. its distance to sensor devices. The downside of this approach is that it increases the size of the maps and the observation space even further, therefore requiring larger networks with more trainable parameters. 

The second map processing step is to present the centered map as a compressed global and uncompressed but cropped local map as previously evaluated in \cite{Theile2021}. In path planning, as distant features lead to general direction decisions while close features lead to immediate actions such as collision avoidance, the level of detail passed to the agent for distant objects can be less than for close objects. The advantage is that the compression of the global map reduces the necessary neural network size considerably. 

This reduction in network size directly translates to a reduction in computational load. Table \ref{tab:flops} shows the number of floating point operations needed for each of the two maps under different map processing regimes as given by the TensorFlow graph profiler. Only centering increases the computational load considerably, as explained in \cite{Bayerlein2020}, while global-local map processing offsets the increase and reduces floating point operations considerably. Considering that modern embedded processors operate in the region of giga floating point operations, it seems realistic that the required processing can be carried out even on small and energy-limited UAVs. The mathematical descriptions of the map processing functions and the observation space are detailed in the following.

\begin{table}[h]
    \centering
    \renewcommand{\arraystretch}{1.2}
    \begin{tabular*}{0.99\columnwidth}{cccc}
        \toprule[1.5pt]
         Map & No Processing & Centering & Centering + Global-Local  \\ 
        \midrule
         Manhattan32 & 15 & 80 & 7.7 \\
         Urban50 & 45 & 217 & 6.5\\
         \bottomrule[1.5pt]
    \end{tabular*}
    \vspace{6pt}
    \caption{Million floating point operations (MFLOPs) needed for inference of the networks based on map-processing.}
    \label{tab:flops}
\end{table}

\subsection{Map-Processing}
\label{subsec:input_space}
For ease of exposition, we introduce the 2D projections of the UAV and IoT device positions on the ground, $\tilde{\mathbf{u}}_k \in \mathbb{N}^2$ and $\tilde{\mathbf{p}}_k \in \mathbb{N}^2$ respectively, given by
\begin{equation}
  \tilde{\mathbf{u}}_k = \left\lfloor\begin{pmatrix}
\frac{1}{c} & 0 & 0\\
0 & \frac{1}{c} & 0
\end{pmatrix} \mathbf{u}_k\right\rceil, \quad   
\tilde{\mathbf{p}}_i = \left\lfloor\begin{pmatrix}
\frac{1}{c} & 0 & 0\\
0 & \frac{1}{c} & 0
\end{pmatrix} \mathbf{p}_i\right\rceil
\end{equation}
rounded to integer grid coordinates.

\subsubsection{Mapping}
The centering and global-local mapping algorithms are based on map-layer representations of the state space. To represent any state with a spatial aspect given by a position and a corresponding value as a map-layer, we define a general mapping function 
\begin{equation}
f_\text{mapping} : \mathbb{N}^{Q\times 2} \times \mathbb{R}^Q \mapsto \mathbb{R}^{M\times M}.
\end{equation}
In this function, a map layer $\mathbf{A}\in\mathbb{R}^{M\times M}$ is defined as
\begin{equation}
    \mathbf{A} = f_\text{mapping}(\{\tilde{\mathbf{p}}_q\}, \{v_q\}), 
\end{equation}
with a set of grid coordinates $\{\tilde{\mathbf{p}}_q\}$ and a set of corresponding values $ \{v_q\}$. The elements of $ \mathbf{A}$ are given through
\begin{equation}
    a_{\tilde{p}_{q,0}, \tilde{p}_{q,1}} = v_q, \quad \forall q \in [0,...,Q-1] 
\end{equation}
or $0$ if the index is not in the grid coordinates. With this general function, we define the map-layers
\begin{subequations}
\begin{align}
    \mathbf{D}(t) = &f_\text{mapping}(\{\tilde{\mathbf{u}}_k\}, \{D_k(t)\})\\
    \mathbf{B}(t) = &f_\text{mapping}(\{\tilde{\mathbf{p}}_i(t)\}, \{b_i(t)\})\\
    \Phi(t) = &f_\text{mapping}(\{\tilde{\mathbf{p}}_i(t)\}, \{\phi_i(t)\})
\end{align}
\end{subequations}
for device data, UAV flying times, and UAV operational status respectively. If the map-layers are of same type they can be stacked to form a tensor of $\mathbb{R}^{M\times M\times n}$ for ease of representation.

\subsubsection{Map Centering}
\begin{figure}
    \centering
    \begin{subfigure}{0.49\columnwidth}
        \centering
        \includegraphics[width=\textwidth]{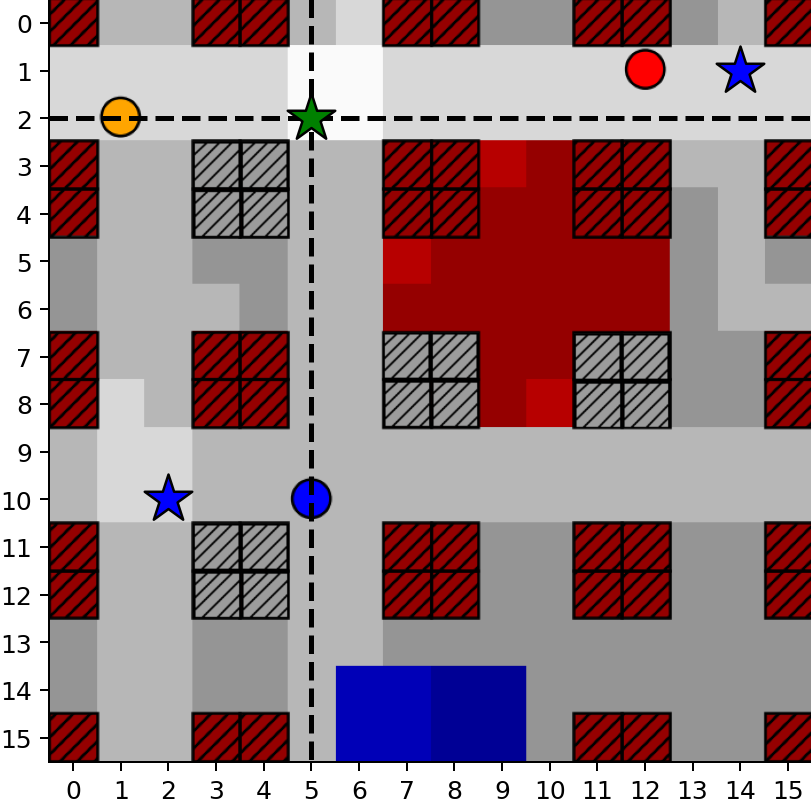}
    \end{subfigure}\hspace{5pt}%
    \begin{subfigure}{0.49\columnwidth}
        \centering
        \includegraphics[width=\textwidth]{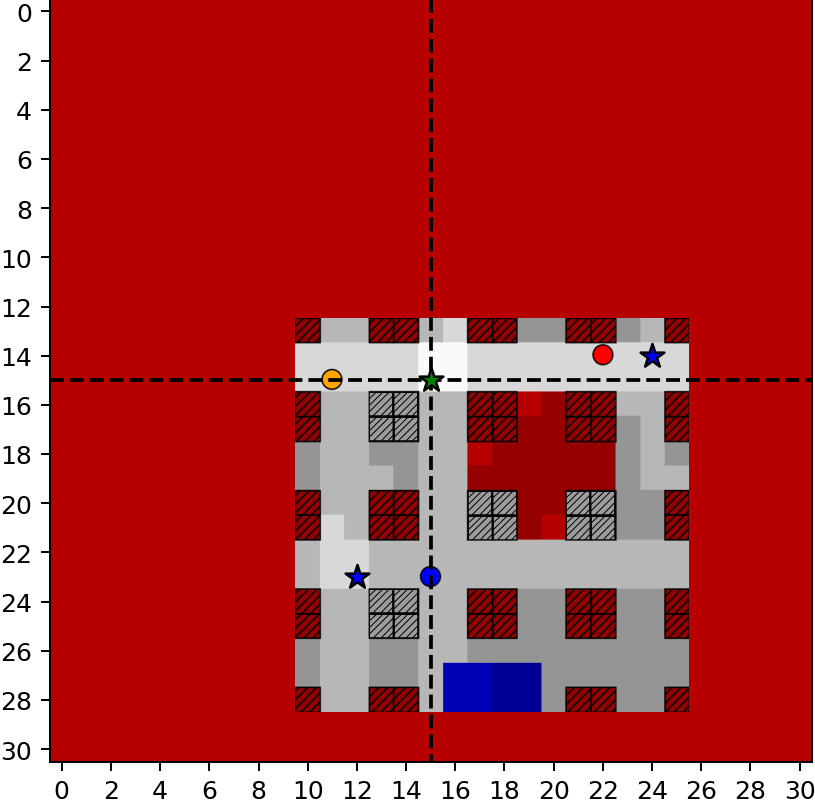}
    \end{subfigure}\\\hspace{0.015\columnwidth}
    \begin{subfigure}{0.46\columnwidth}
    $\underbrace{
        \hphantom{\includegraphics[width=\textwidth]{graphics/not_centered_new.png}}
        }_{\Large M}$
        \caption{Non-centered input map}
    \end{subfigure}\hspace{0.05\columnwidth}%
    \begin{subfigure}{0.46\columnwidth}
    $\underbrace{
        \hphantom{\includegraphics[width=\textwidth]{graphics/not_centered_new.png}}
        }_{\Large M_c}$
        \caption{Centered input map}
    \end{subfigure}
    \caption{Comparison of non-centered and centered input maps, with UAV position represented by the green star and the intersection of the dashed lines.}
    \label{fig:centering}
\end{figure}

Given a tensor $\mathbf{A} \in \mathbb{R}^{M\times M\times n}$ describing the map-layers, a centered tensor $\mathbf{B} \in \mathbb{R}^{M_c\times M_c\times n}$ with $M_c = 2M-1$ is defined through 
\begin{equation}
    \mathbf{B} = f_\text{center}(\mathbf{A}, \tilde{\mathbf{p}}, \mathbf{x}_{\text{pad}}), 
\end{equation}
with the centering function defined as
\begin{equation}
    f_\text{center}: \mathbb{R}^{M\times M\times n} \times \mathbb{N}^2\times \mathbb{R}^n \mapsto \mathbb{R}^{M_c\times M_c\times n}.
\end{equation}
The elements of $\mathbf{B}$ with respect to the elements of $\mathbf{A}$ are defined as
\begin{equation}
\mathbf{b}_{i,j} = 
    \begin{cases}
    \mathbf{a}_{i+\tilde{p}_0 - M + 1, j+\tilde{p}_1 - M + 1}, &\phantom{\land~}M \leq i+\tilde{p}_0+1 < 2M\\
    & \land~ M \leq j+\tilde{p}_1+1 < 2M\\
    \mathbf{x}_{\text{pad}}, & \text{otherwise},
    \end{cases}
\end{equation}
effectively padding the map layers of $\mathbf{A}$ with the padding value $\mathbf{x}_\text{pad}$. Note that $\mathbf{a}_{i,j}$, $\mathbf{b}_{i,j}$, and $\mathbf{x}_\text{pad}$ are vector valued of dimension $\mathbb{R}^n$. An illustration of the centering on a $16\times 16$ map ($M = 16$, $M_c = 31$) can be seen in Figure \ref{fig:centering} with the legend in Table \ref{table:legend}.

\subsubsection{Global-Local Map}
The tensor $\mathbf{B} \in \mathbb{R}^{M_c\times M_c\times n}$ resulting from the map centering function is processed in two ways. The first is creating a local map according to
\begin{equation}
    \mathbf{X} = f_\text{local}(\mathbf{B},l)
\end{equation}
with the local map function defined by
\begin{equation}
    f_\text{local}: \mathbb{R}^{M_c\times M_c\times n} \times \mathbb{N} \mapsto \mathbb{R}^{l\times l\times n}.
\end{equation}
The elements of $\mathbf{X}$ with respect to the elements of $\mathbf{B}$ are defined as
\begin{equation}
\mathbf{x}_{i,j} = \mathbf{b}_{i+M-\lceil\frac{l}{2} \rceil, j+M-\lceil\frac{l}{2} \rceil}
\end{equation}
This operation is effectively a central crop of size $l\times l$.

The second processing creates a global map according to
\begin{equation}
    \mathbf{Y} = f_\text{global}(\mathbf{B},g)
\end{equation}
with the global map function
\begin{equation}
    f_\text{global}: \mathbb{R}^{M_c\times M_c\times n} \times \mathbb{N} \mapsto \mathbb{R}^{\lfloor\frac{M_c}{g} \rfloor\times \lfloor\frac{M_c}{g} \rfloor\times n}
\end{equation}
The elements of $\mathbf{Y}$ with respect to the elements of $\mathbf{B}$ are defined as
\begin{equation}
\mathbf{y}_{i,j} = \frac{1}{g^2} \sum_{u=0}^{g-1}\sum_{v=0}^{g-1} \mathbf{b}_{gi+u, gj+v}
\end{equation}
This operation is equal to an average pooling operation with pooling cell size $g$. 

The functions $f_\text{local}$ and $f_\text{global}$ are parameterized through $l$ and $g$, respectively. Increasing $l$ increases the size of the local map, whereas increasing $g$ increases the size of the average pooling cells, therefore decreasing the size of the global map.

\subsection{Observation Space}
Using the map processing functions, the observation space can be defined. The observation space $\Omega$, which is the input space to the agent, is given as
\begin{equation*}
    \Omega = \underbrace{\Omega_{l}}_{\substack{\text{Local}\\\text{Map}}} \times \underbrace{\Omega_g}_{\substack{\text{Global}\\\text{Map}}} \times \underbrace{\mathbb{N}}_{\substack{\text{Flying}\\\text{Time}}}
\end{equation*}
containing the local map
\begin{equation*}
    \Omega_l = \mathbb{B}^{l\times l\times 3}\times \mathbb{R}^{l\times l}\times \mathbb{N}^{l\times l}\times \mathbb{B}^{l\times l}
\end{equation*}
and the global map
\begin{equation*}
    \Omega_g = \mathbb{R}^{\bar{g}\times \bar{g}\times 3}\times \mathbb{R}^{\bar{g}\times \bar{g}}\times \mathbb{R}^{\bar{g}\times \bar{g}}\times \mathbb{R}^{\bar{g}\times \bar{g}}.
\end{equation*}
with $\bar{g}=\lfloor\frac{M_c}{g}\rfloor$. Note that the compression of the global map through average pooling transforms all map layers into $\mathbb{R}$. Observations $o_i(t) \in \Omega$ are defined through the tuple
\begin{align}
    o_i(t) = (&\mathbf{M}_{l,i}(t),\mathbf{D}_{l,i}(t),\mathbf{B}_{l,i}(t),\Phi_{l,i}(t),\nonumber\\
    &\mathbf{M}_{g,i}(t), \mathbf{D}_{g,i}(t), \mathbf{B}_{g,i}(t),\Phi_{g,i}(t), b_i(t)).
\end{align}
In one observation tuple, $\mathbf{M}_{l,i}(t)$ is the local observation of agent $i$ of the environment, $\mathbf{D}_{l,i}(t)$ is the local observation of the data to be collected, $\mathbf{B}_{l,i}(t)$ is the local observation of the remaining flying time of all agents, and $\Phi_{l,i}(t)$ is the local observation of the operational status of the agents. $\mathbf{M}_{g,i}(t)$, $\mathbf{D}_{g,i}(t)$, $\mathbf{B}_{g,i}(t)$, and $\Phi_{g,i}(t)$ are the respective global observations. $b_i(t)$ is the remaining flying time of agent $i$, which is equal to the one in the state space. Note that the environment map's local and global observations are dependent on time, as they are centered around the UAV's time-dependent position. Additionally, it should be noted that the remaining flying time of agent $i$ is given in the center of $\mathbf{B}_{l,i}(t)$ and additionally as a scalar $b_i(t)$. This redundancy in representation helps the agent to interpret the remaining flying time.

Consequently, the complete mapping from state to observation space is given by
\begin{equation}
    \mathcal{O} : \mathcal{S}\times\mathcal{I}\mapsto \Omega
\end{equation}
in which the elements of $o_i(t)$ are defined as follows:
\begin{subequations}\label{eq:observation}
\begin{align}
    \mathbf{M}_{l,i}(t) =& f_\text{local}(f_\text{center}(\mathbf{M},\mathbf{p}_i(t),[0,1,1]^\text{T}), l)\\
    \mathbf{D}_{l,i}(t) =& f_\text{local}(f_\text{center}(\mathbf{D}(t),\mathbf{p}_i(t),0), l)\\
    \mathbf{B}_{l,i}(t) =& f_\text{local}(f_\text{center}(\mathbf{B}(t),\mathbf{p}_i(t),0), l)\\
    \Phi_{l,i}(t) =& f_\text{local}(f_\text{center}(\Phi(t),\mathbf{p}_i(t),0), l)\\
    \mathbf{M}_{g,i}(t) =& f_\text{global}(f_\text{center}(\mathbf{M},\mathbf{p}_i(t),[0,1,1]^\text{T}), g)\\
    \mathbf{D}_{g,i}(t) =& f_\text{global}(f_\text{center}(\mathbf{D}(t),\mathbf{p}_i(t),0), g)\\
    \mathbf{B}_{g,i}(t) =& f_\text{global}(f_\text{center}(\mathbf{B}(t),\mathbf{p}_i(t),0), g)\\
    \Phi_{g,i}(t) =& f_\text{global}(f_\text{center}(\Phi(t),\mathbf{p}_i(t),0), g)
\end{align}
\end{subequations}

By passing the observation space $\Omega$ into the agent instead of the state space $\mathcal{S}$ as done in the previous approaches \cite{Theile2020} and \cite{Bayerlein2020}, the presented path planning problem is artificially converted into a partially observable MDP. Partial observability is a consequence of the restricted size of the local map and the compression of the global map. However, as shown in \cite{Theile2021}, partial observability does not render the problem infeasible, even for a memory-less agent. Instead, the compression greatly reduces the neural network's size, leading to a significant reduction in training time.

\section{Multi-Agent Reinforcement Learning (MARL)}
\label{sec:marl}

\subsection{Q-Learning}

Q-learning is a model-free RL method \cite{Sutton2018} where a cycle of interaction between one or multiple agents and the environment enables the agents to learn and optimize a behavior, i.e. the agents observe state $s_t \in \mathcal{S}$ and each performs an action $a_t \in \mathcal{A}$ at time $t$ and the environment subsequently assigns a reward $r(s_t, a_t) \in \mathbb{R}$ to the agents. The cycle restarts with the propagation of the agents to the next state $s_{t+1}$. The agents' goal is to learn a behavior rule, referred to as a policy that maximizes their reward. A probabilistic policy $\pi(a|s)$ is a distribution over actions given the state such that $\pi : \mathcal{S}\times\mathcal{A}\rightarrow\mathbb{R}$. In the deterministic case, it reduces to $\pi(s)$ such that $\pi : \mathcal{S}\rightarrow\mathcal{A}$.

Q-learning is based on iteratively improving the state-action value function or Q-function to guide and evaluate the process of learning a policy $\pi$. It is given as
\begin{equation}
    Q^\pi(s,a) = \mathbb{E}_{\pi} \left[G_t | s_t = s, a_t = a\right]
    \label{eq:q}
\end{equation}
and represents an expectation of the discounted cumulative return $G_t$  from the current state $s_t$ up to a terminal state at time $T$ given by
\begin{equation}
    G_t = \sum_{k=t}^T \gamma^{k-t} r(s_k,a_k)
\end{equation}
with $\gamma \in [0, 1]$ being the discount factor, balancing the importance of immediate and future rewards. For the ease of exposition, $s_t$ and $a_t$ are abbreviated to $s$ and $a$, while $s_{t+1}$ and $a_{t+1}$ are abbreviated to $s^\prime$ and $a^\prime$ in the following.

\begin{figure*}[t]
    \centering
    \includegraphics[width=0.86\textwidth]{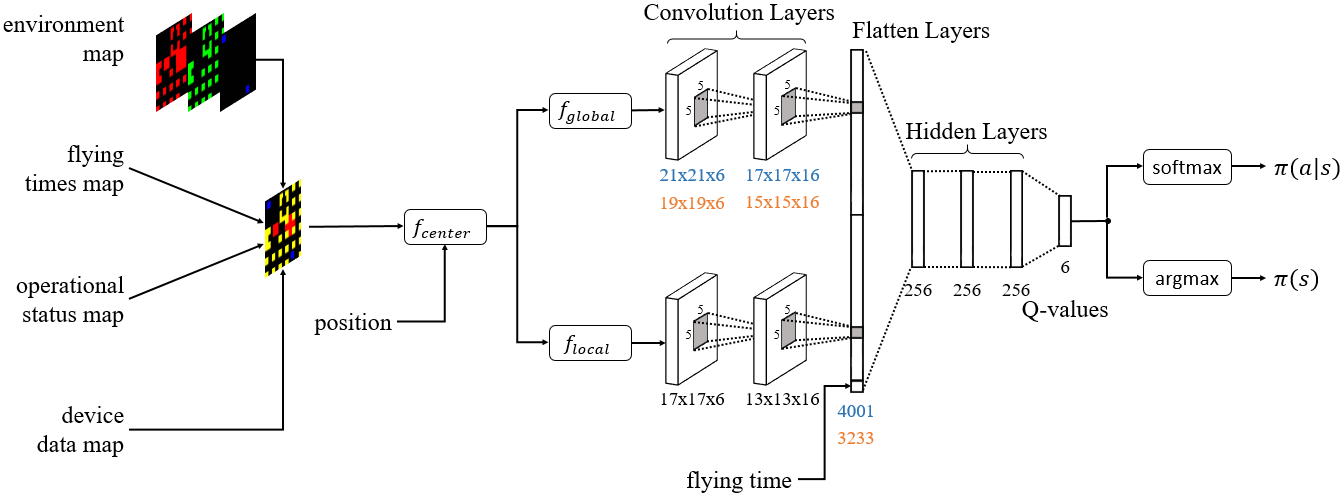}
    \caption{DQN architecture with map centering and global and local map processing. Layer sizes are shown in in blue for the smaller 'Manhattan32' scenario and orange for the larger 'Urban50' scenario.}
    \label{fig:network}
    \vspace{15pt}
\end{figure*}

\subsection{Double Deep Q-learning and Combined Experience Replay}

As demonstrated in \cite{Bayerlein2018spawc}, representing the Q-function \eqref{eq:q} as a table of values is not efficient in the large state and action spaces of UAV trajectory planning. Instead, a deep Q-network (DQN) parameterizing the Q-function with the parameter vector $\theta$ can be trained to minimize the expected temporal difference (TD) error. While a neural network is significantly more data efficient compared to a Q-table due to its ability to generalize, the \mbox{\textit{deadly triad} \cite{Sutton2018}} of function approximation, bootstrapping and off-policy training can make its training unstable and cause divergence.

Mnih \textit{et al.} \cite{Mnih2015} applied stabilizing techniques to the DQN training process, such as experience replay, reducing correlations in the sequence of training data. New experiences made by the agent, represented by quadruples of $(s, a, r, s^\prime)$, are stored in the replay memory $\mathcal{D}$. During training, a minibatch of size $m$ is sampled uniformly from $\mathcal{D}$ and used to compute the loss. The size of the replay memory $\mathcal{|\mathcal{D}|}$ was shown to be an essential hyperparameter for the agent's learning performance and typically must be carefully tuned for different tasks or scenarios. Zhang and Sutton \cite{Zhang2017} proposed combined experience replay as a remedy for this sensitivity with very low computational complexity $\mathcal{O}(1)$. In this extension to the replay memory method, only $m-1$ samples of the minibatch are sampled from memory, and the latest experience the agent made is always added. This corrected minibatch is then used to train the agent. Therefore, all new transitions influence the agent immediately, making the agent less sensitive to the selection of the replay buffer size in our approach.

In addition to experience replay, Mnih \textit{et al.} used a separate target network for the estimation of the next maximum Q-value, giving the loss as
\begin{equation}
    L^{\text{DQN}}(\theta) = \mathbb{E}_{s,a,s^\prime \sim \mathcal{D}}[(Q_\theta(s,a) - Y^{\text{DQN}}(s,a,s^\prime))^2]
\end{equation}
with target value
\begin{equation}
    Y^{\text{DQN}}(s,a,s^\prime) = r(s,a) + \gamma \max_{a^\prime}Q_{\bar{\theta}}\left(s^\prime, a^\prime\right).
\end{equation}
$\bar{\theta}$ represents the parameters of the target network. The parameters of the target network $\bar{\theta}$ can either be updated as a periodic hard copy of $\theta$ or as in our approach with a soft update
\begin{equation}
    \bar{\theta} \leftarrow (1-\tau)\bar{\theta} + \tau\theta
    \label{eq:soft_update}
\end{equation}
after each update of $\theta$. $\tau \in [0,1]$ is the update factor determining the adaptation pace.

Further improvements to the training process were suggested in \cite{VanHasselt2016}, resulting in the inception of double deep Q-networks (DDQNs). With the application of this extension, we avoid the overestimation of action values under certain conditions in standard DQN and arrive at the loss function for our network given by
\begin{equation}
    L^{\text{DDQN}}(\theta) = \mathbb{E}_{s,a,s^\prime \sim \mathcal{D}}[(Q_\theta(s,a) - Y(s,a,s^\prime))^2]
    \label{eq:loss}
\end{equation}
where the target value is given by 
\begin{equation}
    Y^{\text{DDQN}}(s,a,s^\prime) = r(s,a) + \gamma Q_{\bar{\theta}}(s^\prime, \argmax_{a^\prime}Q_{\theta}(s^\prime, a^\prime)).
    \label{eq:target}
\end{equation}

\subsection{Multi-agent Q-learning}
\label{subsec:multi_agent}

The original table-based Q-learning algorithm was extended to the cooperative multi-agent setting by Claus and Boutilier in 1998 \cite{Claus1998}. Without changing the underlying principle, it can also be applied to DDQN-based multi-agent cooperation. With the taxonomy from \cite{Stone2000}, our agents can be classified as homogeneous and non-communicating. Homogeneity is a consequence of deploying a team of identical UAVs with the same internal structure, domain knowledge, and identical action spaces. Non-communication is to be interpreted in a multi-agent system sense, i.e. that the agents can not coordinate their actions or choose what to communicate. However, as they all perceive state information that includes other UAVs' positions, in a practical sense, position information would most likely be communicated via the command and control links of the UAVs, that especially autonomous UAVs would have to maintain for regulatory purposes in any case.

The best way to describe our learning approach is by decentralized deployment or execution with centralized training. As DDQN learning requires an extensive experience database to train the neural networks on, it is reasonable to assume that the experiences made by independently acting agents can be centrally pooled throughout the training phase. After training has concluded, the control systems are individually deployed to the distributed drone agents. The rationale behind this concept is that we investigate a team of homogeneous UAVs with identical capabilities and tasks, therefore all experiences are useful for the training of all agents. In a real-world deployment of a team of quadcopter UAVs, all UAVs would be required to regularly return to a charging station, as flying time remains strongly limited by available on-board battery capacity. While being recharged, the UAVs would upload their experience data to a central server with larger memory and computation resources.

Our setting can not be characterized as fully cooperative as our agents do not share a common reward ~\cite{ Zhang2019}. Instead, each agent has an individual but identical reward function. As the main component of the reward function is based on the jointly collected data from the IoT devices described in Section \ref{subsec:reward}, they do share a common goal, leading to the classification of our setting as a simple cooperative one.

\subsection{Neural Network Model}
We use a neural network model very similar to the one presented in \cite{Theile2021}. Fig. \ref{fig:network} shows the DQN structure and the map centering and global-local map processing. The map information of the environment, NFZs, obstacles, and start/landing area is stacked with the IoT device map and the map with the other UAVs' flying times and operational status. According to Section \ref{subsec:input_space}, the map is centered on the UAV's position and split into a global and local map. The global and local maps are fed through convolutional layers with ReLU activation and then flattened and concatenated with the scalar input indicating battery content or remaining flight time. After passing through fully connected layers with ReLU activation, the data reaches the last fully-connected layer of size $|\mathcal{A}|$ without activation function, directly representing the Q-values for each action given the input observation. The $\argmax$ of the Q-values, the greedy policy is given by
\begin{equation}
    \pi(s) = \argmax_{a\in\mathcal{A}}Q_\theta(s,a).
    \label{eq:greedy}
\end{equation}
It is deterministic and used when evaluating the agent. During training, the soft-max policy 
\begin{equation}
    \pi(a_i|s) = \frac{\mathrm{e}^{Q_\theta(s,a_i)/\beta}}{\sum_{\forall a_j \in \mathcal{A}}\mathrm{e}^{Q_\theta(s,a_j)/\beta}}
    \label{eq:softmax}
\end{equation}
is used. The temperature parameter $\beta \in \mathbb{R}$ scales the balance of exploration versus exploitation. Hyperparameters are listed in Tab. \ref{table:parameters}.

\begin{table}
\center
\small
\begin{tabular*}{1.0\columnwidth}{cccl}
\toprule[1.5pt]
Parameter & $32 \times 32$ & $50 \times 50$ & Description\\
\midrule
$|\theta|$ & 1,175,302 &  978,694 & trainable parameters\\
$N_{\text{max}}$ & 3,000,000  & 4,000,000& maximum training steps\\
$l$ & 17  & 17 & local map scaling\\
$g$ & 3  & 5 & global map scaling\\
$|\mathcal{D}|$ & \multicolumn{2}{c}{50,000} & replay memory buffer size\\
$m$ & \multicolumn{2}{c}{128} & minibatch size\\
$\tau$ & \multicolumn{2}{c}{0.005} & soft update factor in \eqref{eq:soft_update}\\
$\gamma$ & \multicolumn{2}{c}{0.95} & discount factor in \eqref{eq:target}\\
$\beta$ & \multicolumn{2}{c}{0.1} & temperature parameter \eqref{eq:softmax}\\
\bottomrule[1.5pt]\\
\end{tabular*}
\caption{DDQN Hyperparameters for $32 \times 32$ and $50 \times 50$ maps.}
\label{table:parameters}
\vspace{5pt}
\end{table}

\section{Simulations}
\label{sec:simulation}

\subsection{Simulation Setup}
\label{subsec:sim_setup}

In this work, we aim to provide an algorithm\footnote{The Python code for this work is available under \url{https://github.com/hbayerlein/uav\_data\_harvesting}.} that is able to generalize the learned UAV control policy over a large parameter space that defines the specific data collection scenario. That means that at the start of a new training episode, a set of scenario parameters is sampled randomly from a given range of possible values defining the mission. Then the mission starts and the agents are deployed to collect as much data as possible in the given circumstances. Specifically, we define a new mission through the following randomly varying scenario parameters: 
\begin{itemize}
    \item Number of UAVs deployed;
    \item Number and position of IoT sensor nodes;
    \item Amount of data to be collected from IoT nodes;
    \item Flying time available for UAVs at mission start;
    \item UAV start positions.
\end{itemize}
The exact value ranges from which these parameters are sampled are given in the following Sections \ref{subsec:manhattan32} and \ref{subsec:urban50} depending on the map. We deploy our system on two different maps. In 'Manhattan32', the UAVs fly inside 'urban canyons' through a dense city environment discretized into $32 \times 32$ cells, whereas 'Urban50' is an example of a less dense but larger $50 \times 50$ urban area. Note that we only trained a single agent on each of these maps, meaning that all results discussed in the following are a result of only two trained agents. Generalization over this large parameter space is possible in part due to the learning efficiency benefits from feeding map information centered on the agents' respective positions into the network, as we have described previously in \cite{Bayerlein2020}.

We use the following metrics to evaluate the agents' performance on different maps and under different scenario instances:
\begin{itemize}
    \item \textit{Successful landing}: records whether all agents have landed in time at the end of an episode;
    \item \textit{Collection ratio}: the ratio of total collected data at the end of the mission to the total device data that was available at the beginning of the mission;
    \item \textit{Collection ratio and landed}: the product of \textit{successful landing} and \textit{collection ratio} per episode.
\end{itemize}

Evaluation is challenging as we train a single control policy to generalize over a large scenario parameter space. During training, we evaluate the agents' training progress in a randomly selected scenario every five episodes and form an average over multiple evaluations. A single evaluation could be tainted by unusually easy conditions, e.g. when all devices are placed very close to each other by chance. Therefore, only an average over multiple evaluations can be indicative of the agents' learning progress. As it is computationally infeasible to evaluate the trained system on all possible scenario variations, we perform Monte Carlo analysis on a large number of randomly selected scenario parameter combinations.

Irrespective of the map, the grid cell size is $c=10\si{m}$ and the UAVs fly at a constant altitude of $h=10\si{m}$ over city streets. The UAVs are not allowed to fly over tall buildings, enter NFZs, or leave the respective grid worlds. Each mission time slot $t \in [0,T]$ contains $\lambda = 4$ scheduled communication time slots $n\in [0,N]$. Propagation parameters (see \ref{subsec:comm_model}) are chosen in-line with \cite{channelModelETSI} according to the urban micro scenario with $\alpha_{\text{LoS}} = 2.27$, $\alpha_{\text{NLoS}} = 3.64$, $\sigma_{\text{LoS}}^2 = 2$ and $\sigma_{\text{NLoS}}^2 = 5$. 

Due to the drones flying below or slightly above building height, the wireless channel is characterized by strong LoS/NLoS dependency and shadowing. The shadowing maps used for simulation of the environment were computed using ray tracing from and to the center points of cells based on a variation of Bresenham's line algorithm. Transmission and noise powers are normalized by defining a cell-edge SNR for each map, which describes the SNR between one drone on ground level at the center of the map and an unobstructed IoT device maximally far apart at one of the grid corners. The agents have absolutely no prior knowledge of the shadowing maps or wireless channel characteristics.

\subsection{Training with Map-based vs. Scalar Inputs}
\label{subsec:map_vs_scalar}

\begin{figure}[]
    \captionsetup[subfigure]{aboveskip=-1pt}
    \centering
    \begin{subfigure}{0.85\columnwidth}
        \includegraphics[width=\textwidth]{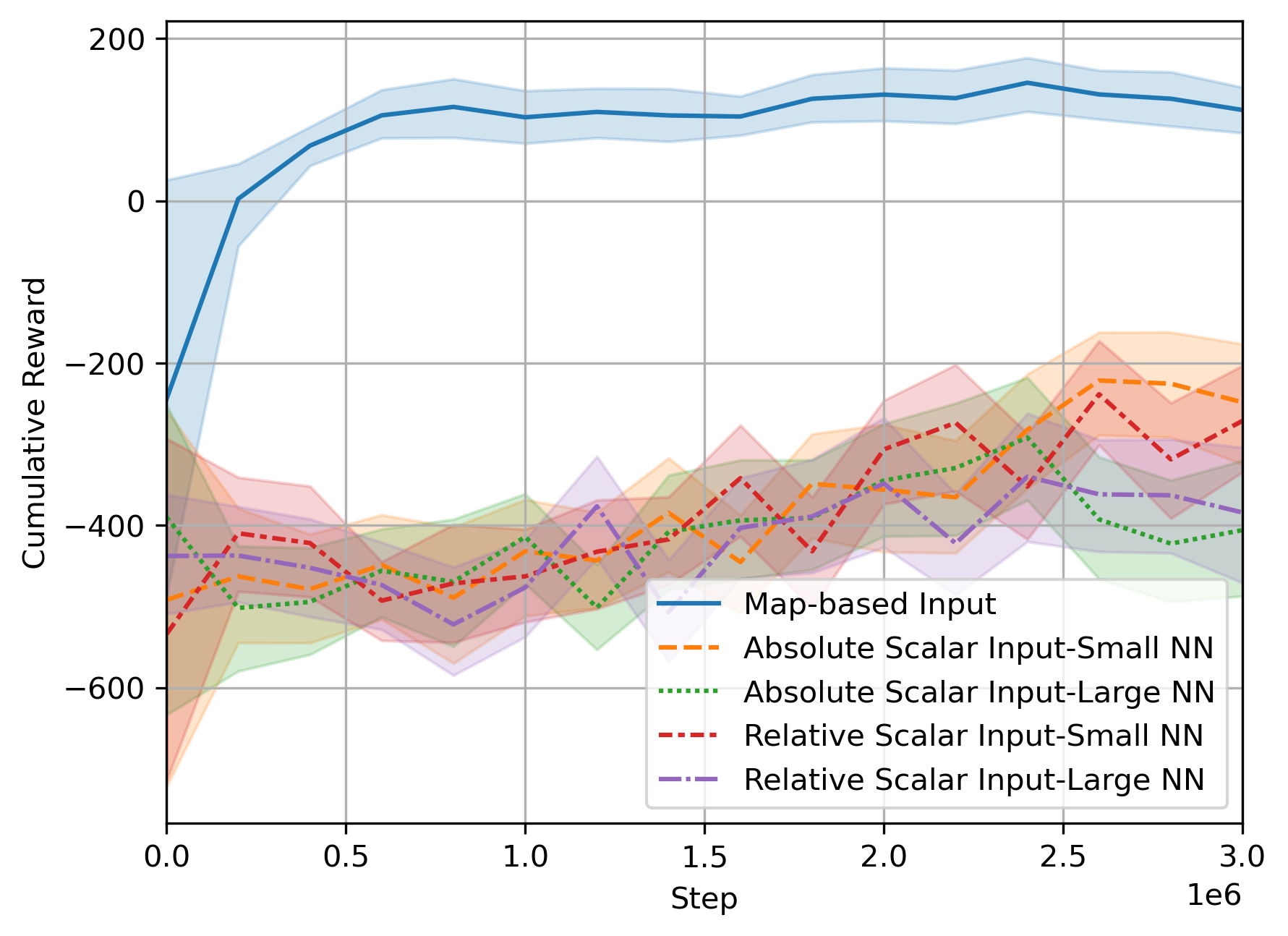}
        \caption{Cumulative reward per episode}
    \end{subfigure}\vspace{5pt}
    \begin{subfigure}{0.826\columnwidth}
    \hspace{1pt}
        \includegraphics[width=\textwidth]{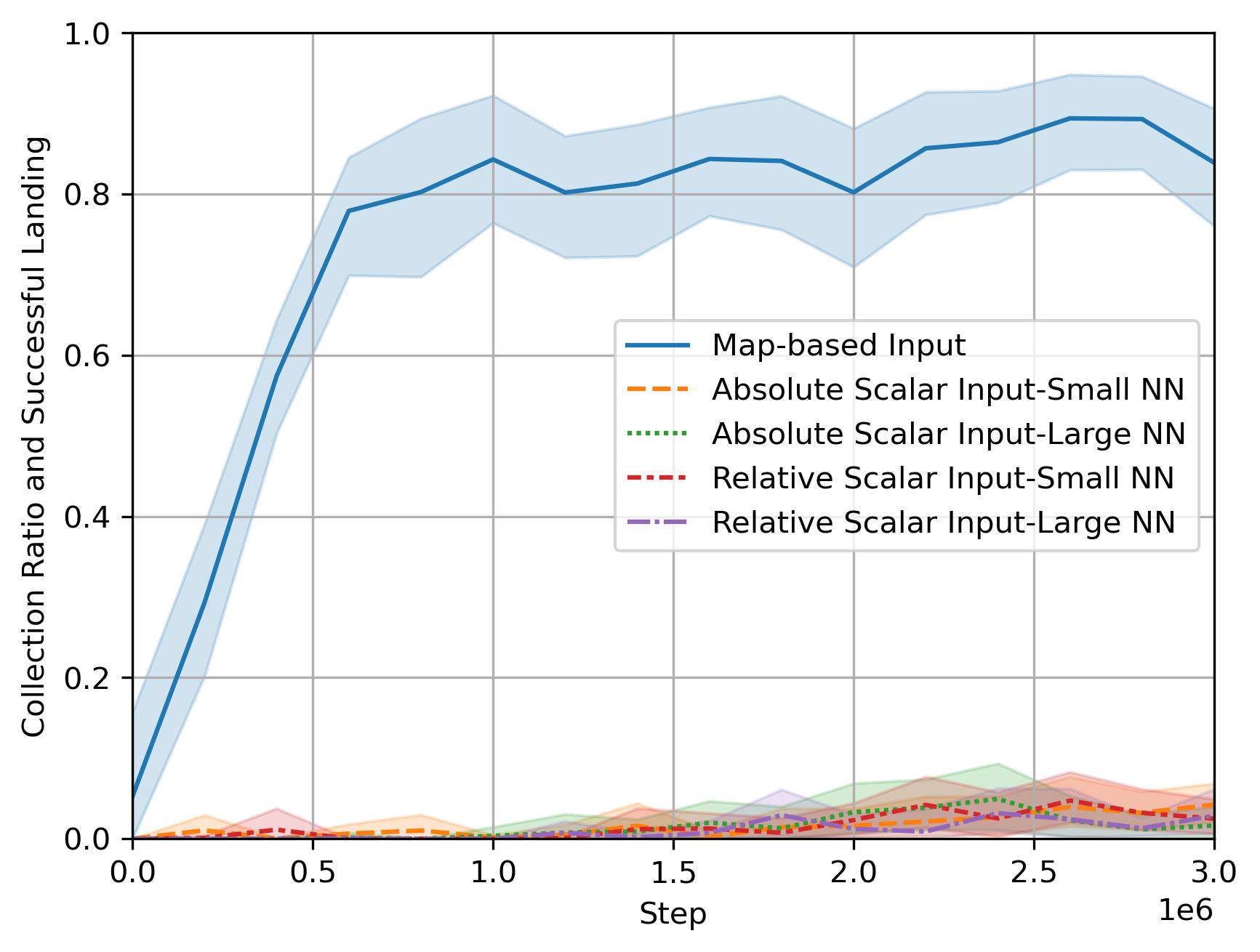}
        \caption{Data collection ratio with successful landing per episode}
    \end{subfigure}
    \caption{Training process comparison between \textit{map-based} DRL path planning and \textit{scalar} input DRL path planning. Scalar inputs to the neural networks (NNs) are either encoded as \textit{absolute} coordinate values or \textit{relative} distances from the respective agent. We compare two different scalar input network architectures with \textit{large} and \textit{small} numbers of trainable parameters. The average and 99\% quantiles are shown with metrics per training episode grouped in bins of $2\mathrm{e}5$ step width. Note that the metrics are plotted over training steps as training episode length is variable.}
    \label{fig:map_scalar}
\end{figure}

In this section, we show that our map-based approach has a good complexity-performance trade-off in comparison to classical scalar input neural network approaches from the literature despite the added complexity through map-processing. To illustrate that it is in fact imperative for training success to feed map information instead of concatenated scalar values as state input to the agent, we extend our previous analysis from \cite{Bayerlein2020} and \cite{Theile2021} by comparing our proposed centered global-local map approach to agents trained only on scalar inputs. This is not an entirely fair comparison as the location of NFZs, buildings, and start/landing zones can not be efficiently represented by scalar inputs and must be therefore learned by the scalar agents through trial and error. However, the comparison illustrates the need for state space representations that are different from the traditional scalar inputs and confirms that scalar agents are not able to solve the multi-UAV path planning problem over the large scenario parameter space presented.
Conversely, the alternative comparison of map-based and scalar agents trained on a \textit{single} data harvesting scenario would not yield meaningful results as our method is specifically designed to generalize over a large variety of scenarios and would require tweaking in exploration behavior and reward balance to find the optimal solution to a single scenario. Note that most of the previous work discussed in section \ref{subsec:related_work} is precisely focused on finding optimal DRL solutions to single scenario instances.

The observation space of the agents trained with concatenated scalar inputs is described by 
\begin{align}
    \mathcal{O}_{\text{scalar}} = &\underbrace{\mathbb{N}^{2}}_{\substack{\text{Ego}\\ \text{Position}}}\times
    \underbrace{ \mathbb{N}}_{\substack{\text{Ego Flying}\\ \text{Time}}}&\Big\}&\text{Ego agent}\nonumber\\
    &\times\underbrace{\mathbb{N}^{I\times 2}}_{\substack{\text{UAV}\\ \text{Positions}}}\times
    \underbrace{ \mathbb{N}^I}_{\substack{\text{Flying}\\ \text{Times}}}\times
    \underbrace{\mathbb{B}^{I}}_{\substack{\text{Operational}\\ \text{Status}}}&\Big\}&\text{Other agents}\label{eq:scalar_obs_space}\\
    &\times\underbrace{\mathbb{N}^{K\times2}}_{\substack{\text{Device}\\ \text{Positions}}}\times
    \underbrace{\mathbb{R}^{K}}_{\substack{\text{Device}\\ \text{Data}}}&\Big\}&\text{Devices}\nonumber
\end{align}
forming the input of the neural network as concatenated scalar values. Since the number of agents and devices is variable, the scalar input size is fixed to the maximum number of agents and devices. The agent and device positions are either represented as \textit{absolute} values in the grid coordinate frame or \textit{relative} as distances from the ego agent. The neural network is either \textit{small}, containing the same number of hidden layers as in Fig. \ref{fig:network}, or \textit{large}, for which the number and size of hidden layers is adapted such that the network has as many trainable parameters as the map-based $32\times32$ agent in Tab. \ref{table:parameters}. 

Fig. \ref{fig:map_scalar} shows the cumulative reward and the collection ratio with successful landing metric over training time on the 'Manhattan32' map for the five different network architectures. It is clear that the scalar agents are not able to effectively adapt to the changing scenario conditions. The \textit{small} neural network agents seem to have a slight edge over the \textit{large} agents, but representing the positions as \textit{absolute} or \textit{relative} does not influence the results.

Referring further to Fig. \ref{fig:map_scalar}, the map-based agent converges to final performance metric levels after the first 20\% of the training steps. However we observed that additional training is needed after that to optimize the trajectories in a more subtle way for flight time efficiency and multi-UAV coordination. The overall training time for the full 3 million training steps was around 40 hours on a 2017 Nvidia Titan Xp GPU.

\subsection{'Manhattan32' Scenario}
\label{subsec:manhattan32}

\begin{figure*}
    \centering
    \begin{subfigure}{0.3\textwidth}
        \centering
        \includegraphics[width=\textwidth]{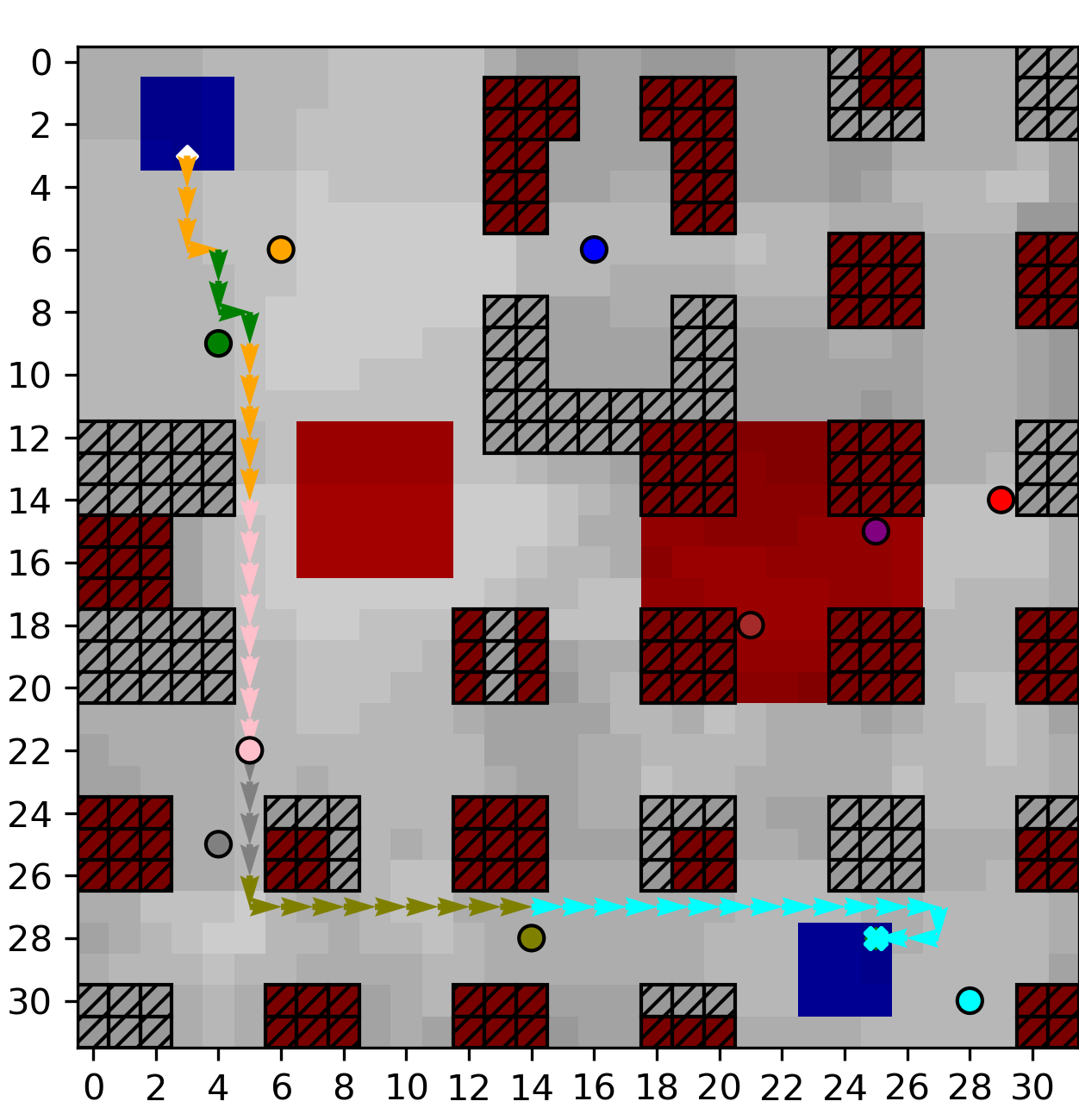}
        \caption{$I=1$ agent, data collection ratio $56.2\%$}
        \label{fig:mh:1}
    \end{subfigure}\hspace{5pt}
    \begin{subfigure}{0.3\textwidth}
        \centering
        \includegraphics[width=\textwidth]{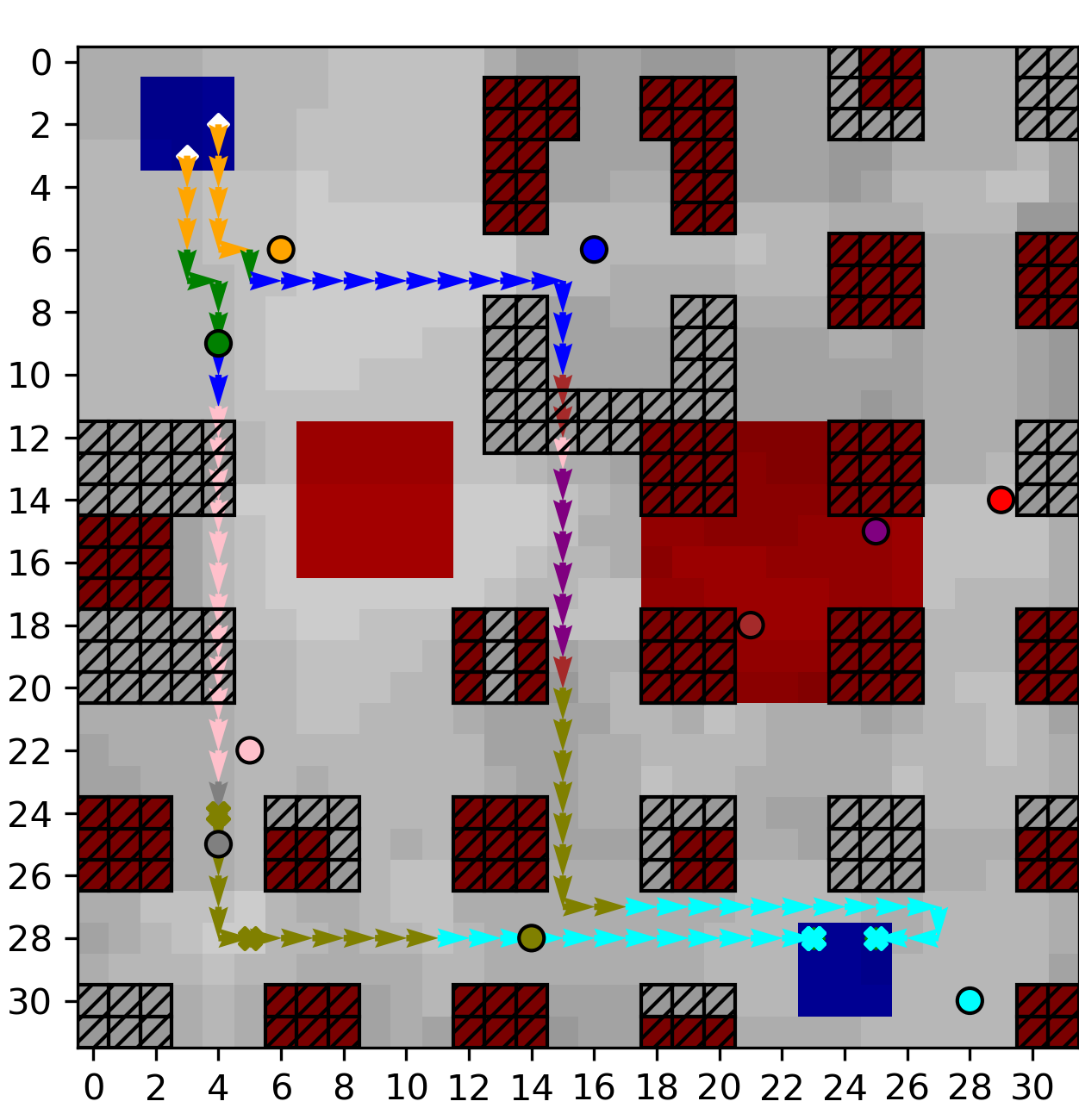}
        \caption{$I=2$ agents, data collection ratio $76.5\%$}
        \label{fig:mh:2}
    \end{subfigure}\hspace{5pt}
    \begin{subfigure}{0.3\textwidth}
        \centering
        \includegraphics[width=\textwidth]{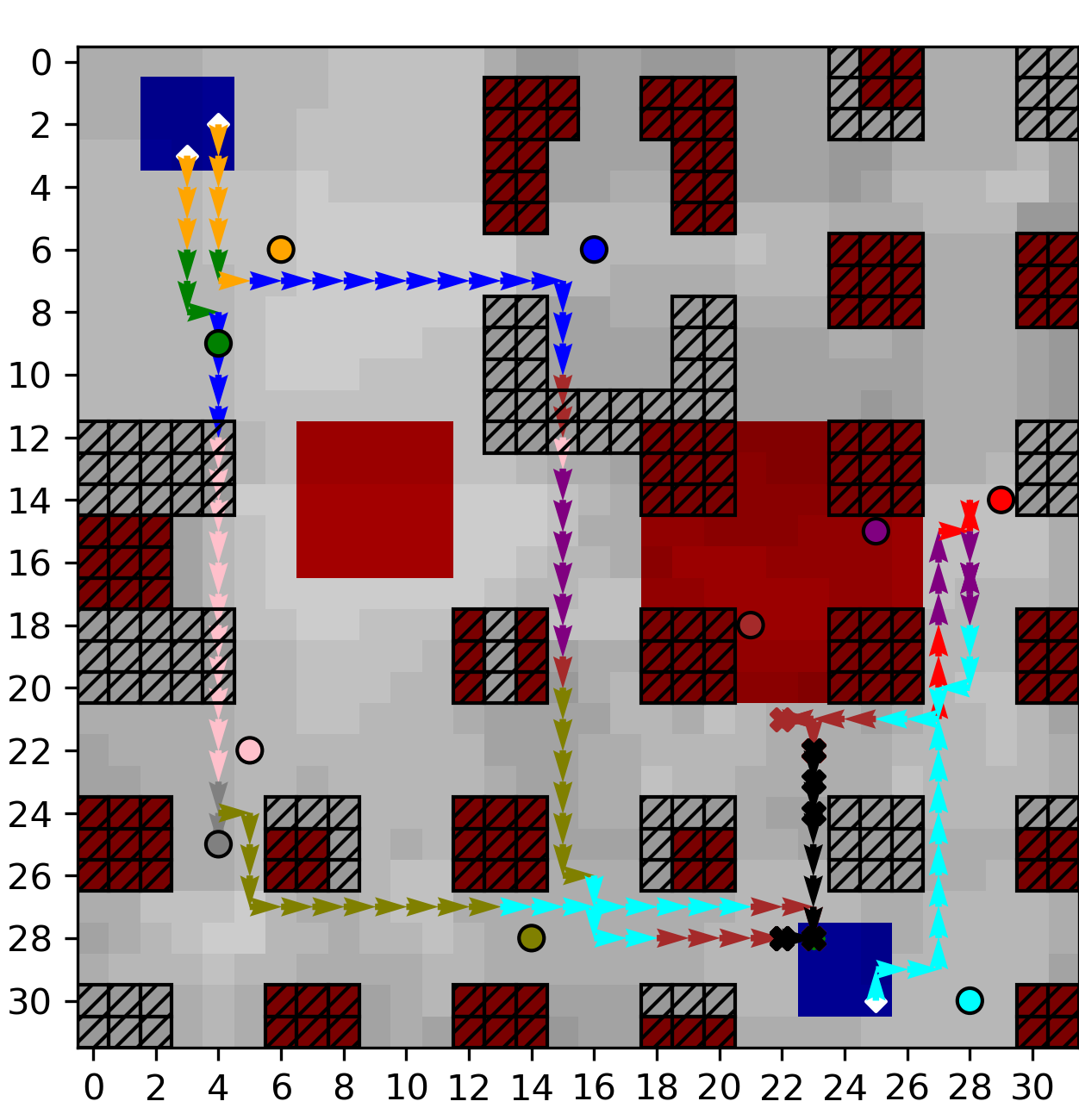}
        \caption{$I=3$ agents, data collection ratio $100\%$}
        \label{fig:mh:3}
    \end{subfigure}
    \caption{Example trajectories for 'Manhattan32' map with $K=10$ IoT devices, all with $D_{k, init}=15$ data units to be picked up and a maximum flying time of $b_0 = 60$ steps. The color of the UAV movement arrows shows with which device the drone is communicating at the time (see legend in Table \ref{table:legend}).}
    \label{fig:mh}
\end{figure*}

The scenario, as shown in Fig. \ref{fig:mh} is defined by a Manhattan-like city structure containing mostly regularly distributed city blocks with streets in between, as well as two NFZ districts and an open space in the upper left corner, divided into $M=32$ cells in each grid direction. This is double the size of the otherwise similarly designed single UAV scenario in \cite{Bayerlein2020}. We are able to solve the larger scenario without increasing network size, thanks to the global-local map approach. The value ranges from which the randomized scenario parameters are chosen as follows: number of deployed UAVs $I \in \{1,2,3\}$, number of IoT sensors $K \in [3,10]$, data volume to be collected $D_{k, init} \in [5.0, 20.0]$ data units per device, maximum flying time $b_0 \in [50, 150]$ steps, and 18 possible starting positions. The IoT device positions are randomized throughout the unoccupied map space.

\begin{table}[]
\center
\small
\begin{tabular*}{0.9\columnwidth}{ccc}
\toprule[1.5pt]
Metric & Manhattan32 & Urban50\\
\midrule
Successful Landing & 99.4\% & 98.8\%\\
Collection Ratio & 88.0\% & 82.1\%\\
Collection Ratio and Landed & 87.5\% & 81.1\%\\
\bottomrule[1.5pt]\\
\end{tabular*}
\caption{Performance metrics averaged over 1000 random scenario Monte Carlo iterations.}
\label{table:metrics}
\end{table}

The performance on both maps is evaluated using Monte Carlo simulations on their respective full range of scenario parameters with overall average performance metrics shown in Table \ref{table:metrics}. Both agents show a similarly high successful landing performance. It is expected that the collection ratio cannot reach 100\% in some scenario instances depending on the randomly assigned maximum flying time, number of deployed UAVs, and IoT device parameters.

In Fig. \ref{fig:mh}, three scenario instances chosen from the random Monte Carlo evaluation for number of deployed UAVs $I \in \{1,2,3\}$ for \ref{fig:mh:1} through \ref{fig:mh:3} illustrate how the path planning adapts to the increasing number of deployed UAVs. All other scenario parameters are kept fixed. It is a fairly complicated scenario with a large number of IoT devices spread out over the whole map, including the brown and purple device inside an NFZ. The agents have no access to the shadowing map and have to deduce shadowing effects from building and device positions.

In Fig. \ref{fig:mh:1}, only one UAV starting in the upper left corner is deployed. Due to its flight time constraint, the agent ignores the blue, red, purple, and brown IoT devices while collecting all data from the other devices on an efficient trajectory to the landing zone in the lower right corner. When a second UAV is deployed in Fig. \ref{fig:mh:2}, the data collection ratio increases to $76.5\%$. While the first UAV's behavior is almost unchanged compared to the single UAV deployment, the second UAV flies to the landing zone in the lower right corner via an alternative trajectory collecting data from the devices the first UAV ignores. With the number of deployed UAVs increased to three (two starting from the upper left and one from the lower right zone) in Fig. \ref{fig:mh:3}, all data can be collected. The second UAV modifies its behavior slightly, accounting for the fact that the third UAV can collect the cyan device's data now. The three UAVs divide the data harvesting task fairly among themselves, leading to full data collection with in-time landing on efficient trajectories while avoiding the NFZs.

\subsection{'Urban50' Scenario}
\label{subsec:urban50}

\begin{figure*}
    \centering
    \begin{subfigure}{0.3\textwidth}
        \centering
        \includegraphics[width=\textwidth]{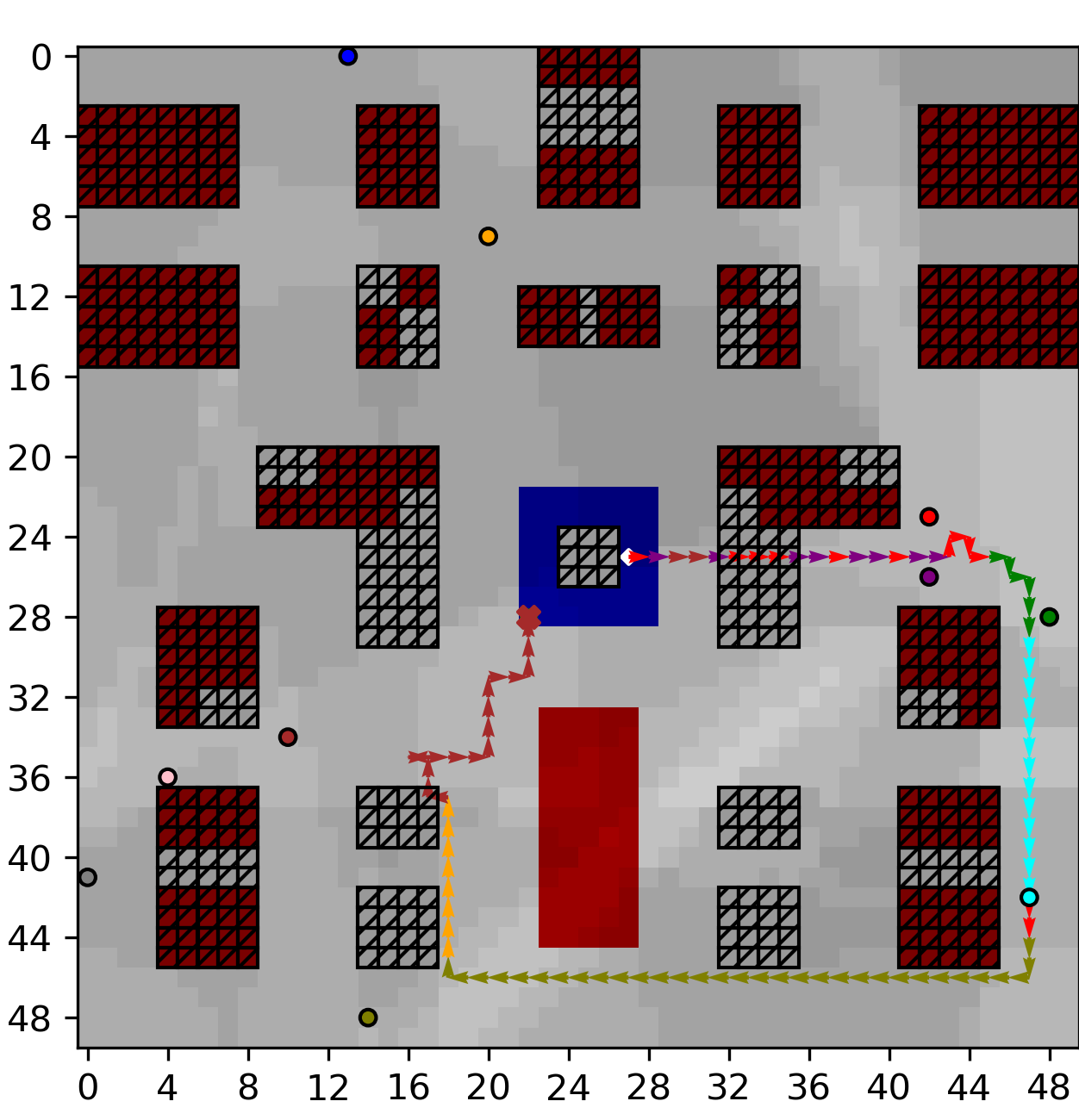}
        \caption{$I=1$ agent, data collection ratio $41.8\%$}
        \label{fig:urban:1}
    \end{subfigure}\hspace{5pt}
    \begin{subfigure}{0.3\textwidth}
        \centering
        \includegraphics[width=\textwidth]{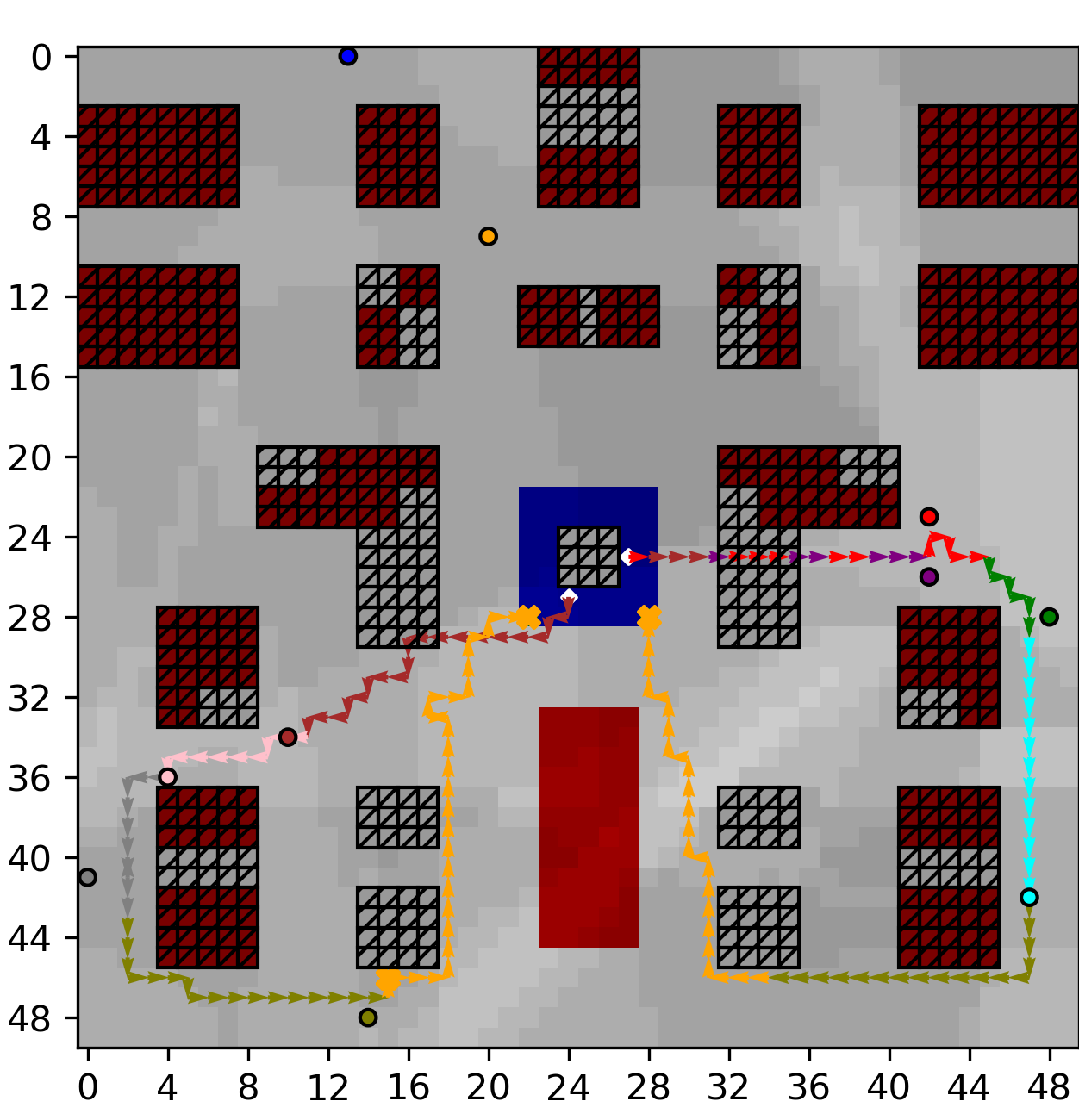}
        \caption{$I=2$ agents, data collection ratio $80.2\%$}
        \label{fig:urban:2}
    \end{subfigure}\hspace{5pt}
    \begin{subfigure}{0.3\textwidth}
        \centering
        \includegraphics[width=\textwidth]{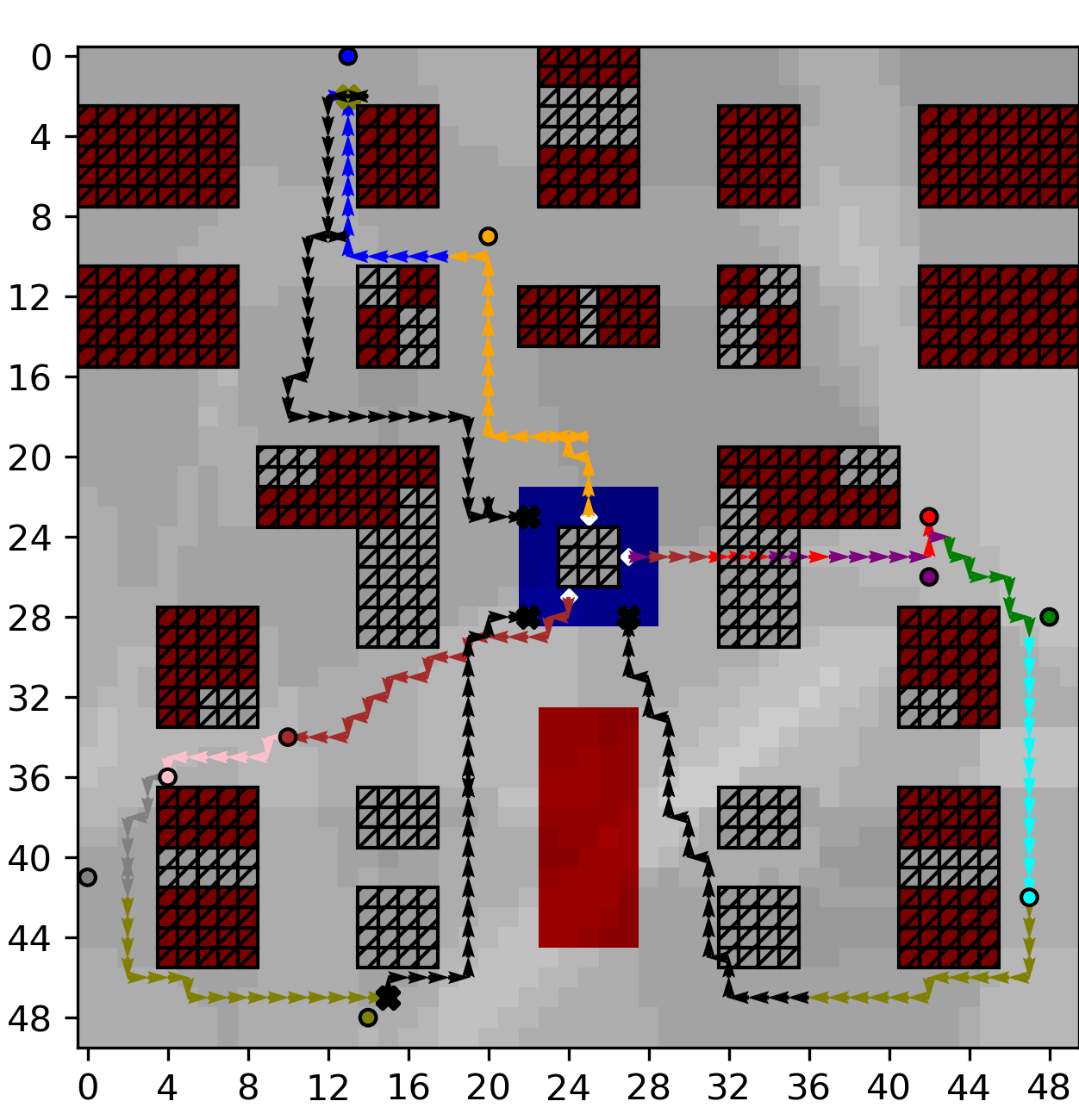}
        \caption{$I=3$ agents, data collection ratio $100.0\%$}
        \label{fig:urban:3}
    \end{subfigure}
    \caption{Example trajectories for 'Urban50' map with $K=10$ IoT devices, all with $D_{k, init}=15$ data units to be picked up and a maximum flying time of $b_0=100$ steps for all UAVs (legend in table \ref{table:legend}).}
    \label{fig:urban}
\end{figure*}

Fig. \ref{fig:urban} shows three example trajectories for UAV counts of $I \in \{1,2,3\}$ for \ref{fig:urban:1} through \ref{fig:urban:3} in the large $50 \times 50$ urban map. The scenario is defined by an urban structure containing irregularly shaped large buildings, city blocks and an NFZ, with the start/landing zone surrounding a building in the center, divided into $M=50$ cells in each grid direction. The map has an order of magnitude more cells than the scenarios in \cite{Bayerlein2020}. The ranges for randomized scenario parameters are chosen as follows: number of deployed UAVs $I \in \{1,2,3\}$, number of IoT sensors $K \in [5,10]$, data volume to be collected $D_{k, init} \in [5.0, 20.0]$ data units, maximum flying time $b_0 \in [100, 200]$ steps, and 40 possible starting positions. The IoT device positions are randomized throughout the unoccupied map space.

Fig. \ref{fig:urban:1} shows a single agent trying to collect as much data as possible during the allocated maximum flying time. The agent focuses on collecting the data from the relatively easily reachable device clusters on the right and lower half before safely landing. With a second UAV assigned to the mission as shown in Fig. \ref{fig:urban:2}, one UAV services the devices on the lower left of the map, while the other one collects data from the devices on the lower right, ignoring the more isolated blue and orange device in the top half of the map. A third UAV makes it possible to divide the map into three sectors and collect all IoT device data, as shown in Fig. \ref{fig:urban:3}.

This map's primary purpose is to showcase the significant advantages in terms of training time efficiency and the required network size from the global-local map approach. Thanks to a higher global map scaling or compression factor $g$ (see Table \ref{table:parameters}), the number of trainable parameters of the network employed in this scenario is even smaller compared to the network used for 'Manhattan32'. A network without a map scaling approach would need to be of size 34,061,446, hence a size that is infeasible to train using reasonable resources. 

\begin{figure*}
    \centering
    \begin{subfigure}[t]{0.47\columnwidth}
        \centering
        \raisebox{8pt}{\includegraphics[width=\textwidth]{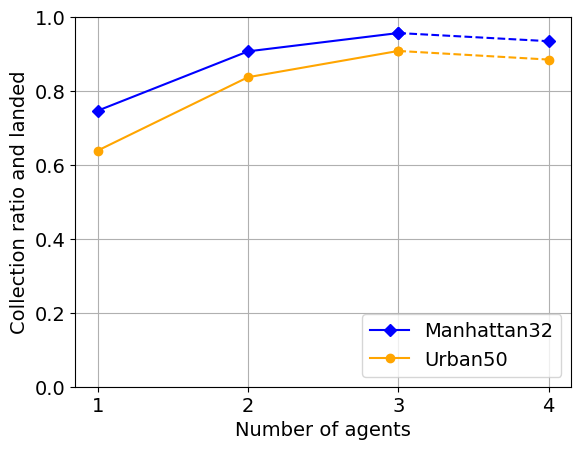}}
        \caption{Number of agents $I \in \{1, 2, 3, 4\}$}.
        \label{fig:stat:1}
    \end{subfigure}\hspace{5pt}
    \begin{subfigure}[t]{0.47\columnwidth}
        \centering
        \raisebox{8pt}{\includegraphics[width=\textwidth]{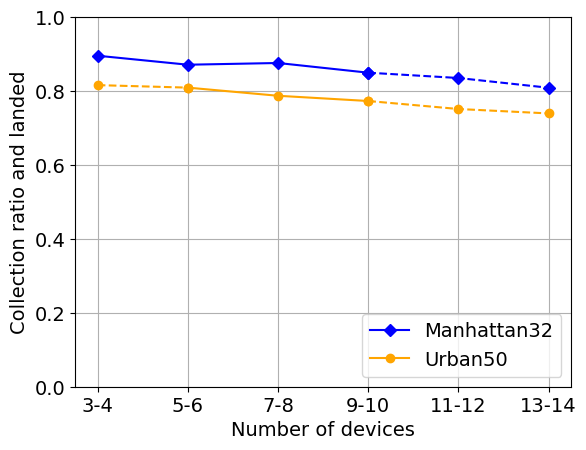}}
        \caption{Number of devices $K \in [3, 14]$ sorted into bins of two.}
        \label{fig:stat:2}
    \end{subfigure}\hspace{5pt}
    \begin{subfigure}[t]{0.47\columnwidth}
        \centering
        \includegraphics[width=\textwidth]{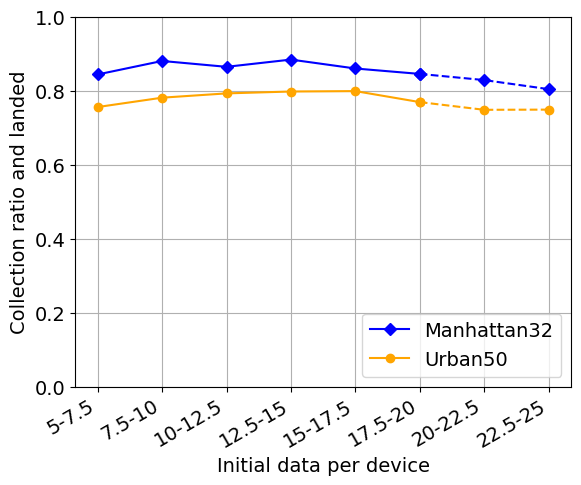}
        \caption{Data to be collected from devices $D_{k, init} \in [5, 25]$ sorted into eight bins.}
        \label{fig:stat:3}
    \end{subfigure}\hspace{5pt}
    \begin{subfigure}[t]{0.47\columnwidth}
        \centering
        \raisebox{8pt}{\includegraphics[width=\textwidth]{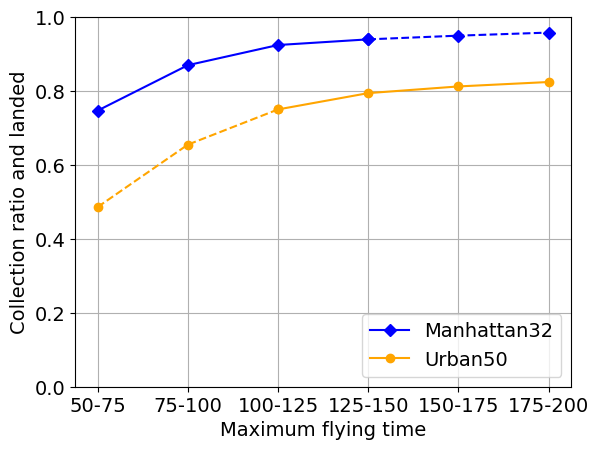}}
        \caption{Maximum flying time sorted into six bins in $b_0 \in [50,200]$}
        \label{fig:stat:4}
    \end{subfigure}
    \caption{Influence of specific scenario parameters on the data collection ratio with successful landing of all agents. Each data point is an average of 500 Monte Carlo iterations over the respective parameter spaces for the 'Manhattan32' and 'Urban50' map. The parameters within the training range are rendered in solid lines and the out-of-distribution parameter evaluation in dashed lines.}
    \label{fig:stat}
\end{figure*}

\subsection{Influence of Scenario Parameters on Performance and System-level Benefits}
\label{subsec:parameters}

An advantage of our approach to learn a generalized path planning policy over various scenario parameters is the possibility to analyze how performance indicators change over a variable parameter space. This makes it possible for an operator to decide on system-level trade-offs, e.g. how many drones to deploy vs. collected data volume. An excellent example that we found for the 'Manhattan32' map was that deploying multiple coordinating drones can trade-off the cost of extra equipment (i.e. the extra drones) for substantially reduced mission time. For instance, it takes twice the flying time ($b_0=150$) for a single UAV to complete the data collection mission that two coordinating UAVs will require ($b_0 = 75$) to conclude successfully. Specifically, that means that for both scenarios the average data collection ratio with in-time successful landing stays at the same performance level of around $88\%$.

We first analyze the performance of the agent in Fig. \ref{fig:stat} within the training range of the scenario parameters (solid lines), then extend the analysis to out-of-distribution scenarios (dashed lines) in the last paragraph of this section. Fig. \ref{fig:stat} shows the influence of single scenario parameters on the average data collection ratio with successful landing of all agents. As already evident from the example trajectories shown previously, Fig. \ref{fig:stat:1} indicates the increase in collection performance when more UAVs are deployed. At the same time, more UAVs lead to increased collision avoidance requirements, as we observed through more safety controller activations in the early training phases. As IoT devices are positioned randomly throughout the unoccupied map space, an increase in devices leads to more complex trajectory requirements and a drop in performance, as depicted in Fig. \ref{fig:stat:2}.

Fig. \ref{fig:stat:3} shows the influence of increasing initial data volume per device on the overall collection performance. It appears that higher initial data volumes per device are beneficial roughly up to the point of $D_{k,init} \in [10,12.5]$ data units, after which flying time constraints force the UAVs to abandon some of the data, and the collection ratio shows a slightly negative trend. An increase in available flying time is clearly beneficial to the collection performance, as indicated in Fig. \ref{fig:stat:4}. However, the effect becomes smaller when most of the data is collected, and the UAVs start to prioritize minimizing overall flight time and safe landing over the collection of the last bits of data.

It is further shown in Fig. \ref{fig:stat} how the agents react to scenario parameters which were not encountered during training. The corresponding values are highlighted with dashed lines. It can be seen that the performance of the agents follows the same trend as in the rest of the data, when increasing the number of devices (Fig. \ref{fig:stat:2}) or initial data per device (Fig. \ref{fig:stat:3}) out of the trained region. When increasing the maximum flying time (Fig. \ref{fig:stat:4}) for the Manhattan32 agents, or decreasing it for Urban50 agents, the collection ratio with successful landing performance, increases or decreases accordingly. Incrementing the number of agents to four (Fig. \ref{fig:stat:1}) reduces the performance slightly. The reason is the decrease in landing performance. However, this is to be expected since the probability of all agents landing decreases with the number of agents. Since the collection ratio is nearly saturated for the scenarios with three agents, the drop in overall landing performance decreases the collection ratio and landed performance. In general, it is evident that the proposed approach cannot only generalize over the whole range of scenario parameters encountered during training but can also extrapolate successfully to out-of-distribution parameters.

\section{Conclusion}
\label{sec:conclusion}

We have introduced a multi-agent reinforcement learning approach that allows us to control a team of cooperative UAVs on a data harvesting mission in a large variety of scenarios without the need for recomputation or retraining when the scenario changes. By leveraging a DDQN with combined experience replay and convolutionally processing dual global-local map information centered on the agents' respective positions, the UAVs are able to find efficient trajectories that balance data collection with safety and navigation constraints without any prior knowledge of the challenging wireless channel characteristics in the urban environments. We have also presented a detailed description of the underlying path planning problem and its translation to a decentralized partially observable Markov decision process. In future work, we will extend the UAVs' action space to altitude and continuous control, requiring an RL algorithm different from Q-learning with a continuous action space and adding height information to the agents' observations space. Moreover, we will investigate if attention-based mechanisms can be used for map processing, assessing their viability with respect to performance and computational requirements in this context. Further improvements in learning efficiency could be achieved when combining our approach with multi-task reinforcement learning or transfer learning \cite{DulacArnold2019}, a step that would bring RL-based autonomous UAV control in the real-world even closer to realization.

\balance
\bibliographystyle{IEEEtran}
\bibliography{bib_ojcoms.bib}

\end{document}